\documentclass[aps,prb,reprint,showpacs]{revtex4-2} % PRL-style format
\usepackage{graphicx} % Required for inserting images
\usepackage{amsmath}  % For mathematical formatting
\usepackage{amssymb}  % For additional symbols
\usepackage[colorlinks=true,linkcolor=blue,urlcolor=blue,citecolor=blue]{hyperref} % for hyperlinks
\usepackage{xr}
\usepackage{tikz}
\usepackage{float}
\usepackage{braket}

\usetikzlibrary{decorations.markings}
\begin{document}

\title{Anharmonic Dephasing in the Electron-Phonon Interaction}
\author{Mingran Kong}
\author{Bartomeu Monserrat}
\affiliation{Department of Materials Science and Metallurgy, University of Cambridge,
27 Charles Babbage Road, Cambridge CB3 0FS, United Kingdom}
\date{\today} % Automatically sets the current date; change if necessary

\begin{abstract}
Electron-phonon coupling has been a central topic in condensed matter physics for decades, and first principles methods have demonstrated remarkable success in quantitatively capturing its role in a wide variety of physical phenomena and materials.
Conventional calculations of electron-phonon coupling typically assume that phonons have infinite lifetimes, but phonons can exhibit finite lifetimes due to anharmonic phonon-phonon interactions.
%, which could affect their ability to mediate coupling.
In this work, we derive an expression for the electron-phonon coupling scattering rates including the effects of anharmonic three-phonon interactions, which lead to phonon dephasing and finite phonon lifetimes.
We also describe a first principles implementation of this anharmonic electron-phonon coupling which can be seamlessly integrated within existing workflows for the evaluation of electron-phonon and phonon-phonon coupling interactions. 
Finally, we present calculations of electron-phonon scattering rates including phonon dephasing in a range of materials, and discuss the different microscopic mechanisms by which anharmonic phonons influence electron-phonon coupling. 
This study establishes the importance of finite phonon lifetimes in the evaluation of electron-phonon coupling, and provides a platform to explore these effects in a wide range of materials and phenomena.
%This study provides a useful reference point and inspires further exploration of electron-phonon interactions beyond the harmonic approximation.
\end{abstract}

\maketitle

\section{Introduction}

The electron-phonon interaction is a cornerstone of condensed matter systems, governing phenomena such as superconductivity, electronic transport, and excited-state dynamics. Over the past few decades, first principles methods have been established as a powerful tool to evaluate numerous material properties derived from electron-phonon coupling. Important examples 
%of discoveries enabled by the efficient and robust calculation of electron-phonon coupling 
include the computational discovery of high-pressure hydride superconductors\,\cite{Duan2014}, the theoretical modeling of electron transport in high-performance thermoelectric materials\,\cite{Wang2012,Coulter2019,He2019}, and the simulation of the optical response of semiconductors\,\cite{Cannuccia2019,Chen2020,Chan2025}.

The conventional framework used to calculate electron-phonon coupling uses density functional theory (DFT) for the description of electrons, the harmonic approximation for the description of phonons calculated within density functional perturbation theory (DFPT)\,\cite{Baroni2001}, 
%or the finite displacement method\,\cite{Alfe2009,Lloyd2015}, 
and a first-order perturbative treatment of the electron-phonon interaction, also calculated within DFPT\,\cite{Baroni2001,Giustino2017}. 
%Within this framework, state-of-the-art calculations use density functional perturbation theory (DFPT), and more recently, also $GW$ perturbation theory ($GW$PT). %provide the foundational methodology, where interpolated coupling matrices feed into self-energy calculations to predict material properties. 
This paradigm has demonstrated remarkable success in the prediction of various quantities derived from electron-phonon coupling across a broad class of materials\,\cite{Ziman2001, Bernardi2016, Kresin2009}.

Recent studies have questioned the conventional paradigm for describing electron-phonon interactions, prompting efforts to extend the theoretical framework. In field-theoretic terms, these extensions can be classified into three categories: electron line corrections, phonon line corrections, and vertex corrections.
The electron line corrections include incorporating many-body effects beyond DFT with approaches such as $GW$ perturbation theory\,\cite{Li2019} and dynamical mean-field theory\,\cite{Abramovitch2023} for a more accurate description of electron-electron interactions; and also include introducing an electron-phonon correction to the electron propagator\,\cite{Lee2020}. 
For the phonon line corrections, variational methods have been developed to construct an effective phonon basis that incorporates anharmonic phonon-phonon interactions\,\cite{Errea2015,Benítez2025}.
Finally, vertex corrections involve introducing coupling strengths beyond linear order\,\cite{Monserrat2013, Lafuente2022PRB, Antonius2015} and exploring higher-order interaction processes\,\cite{Lafuente2022PRB, Lee2020}. Together, these complementary advances refine our understanding of electron-phonon coupling and highlight the need to go beyond conventional approximations to achieve a more comprehensive and predictive theoretical framework.

Nevertheless, all these works describe electron-phonon coupling under the assumption that the phonons involved in the interaction possess infinite lifetimes. This assumption is underpinned by a temporal scale separation between phonon-phonon interactions (characterized by $\tau_{\text{ph-ph}}$) and electron-phonon interactions ($\tau_{\text{e-ph}}$), allowing the treatment of phonons as infinitely long-lived during the electron scattering process. However, there exists some evidence suggesting comparable timescales ($\tau_{\text{ph-ph}} \sim \tau_{\text{e-ph}}$) in numerous materials\,\cite{Liao2015,Wang2017,Yang2021}. This motivates us to explore the role of finite phonon lifetimes in the electron-phonon interaction.

% Inspired by this body of work, we take one step further by introducing dephasing effects to the phonon lines through anharmonic phonon-phonon interactions (PPIs). The standard treatment of EPIs is underpinned by a temporal scale separation between PPIs (characterized by $\tau_{\text{ph-ph}}$) and EPIs ($\tau_{\text{e-ph}}$), allowing the treatment of phonons as infinitely long-lived during the electron scattering process. However, both theoretical and experimental evidence suggest comparable timescales ($\tau_{\text{ph-ph}} \sim \tau_{\text{e-ph}}$) in numerous materials\,\cite{Liao2015,Wang2017,Yang2021}. This motivates us to explore the role of finite phonon lifetimes in the electron-phonon interaction.

In this work, we explore the interplay between finite phonon lifetimes and the electron-phonon interaction, going beyond the traditional separation of phonon and electron time scales. In Sec.\,\ref{sec:theory} we use many-body perturbation theory to derive the lowest order term describing anharmonic dephasing in the electron-phonon interaction, and in Sec.\,\ref{sec:comput} we present the corresponding first principles computational implementation. We then evaluate the anharmonic dephasing in the electron-phonon interactions of three materials: silicon (Si) in Sec.\,\ref{sec:si}, silicon carbide (SiC) in Sec.\,\ref{sec:sic}, and lead telluride (PbTe) in Sec.\,\ref{sec:pbte}. We also highlight a companion work\,\cite{companion} in which we apply this formalism to study electronic transport in metallic magnesium diboride (MgB$_2$), finding that anharmonic dephasing makes a giant contribution to electron-phonon scattering in this compound, which proves key for bringing predictions of electronic conductivity closer to experimental values.

%%%%%%%%%%%%%%%%%%%%%%%%%%%%%%
%%% THEORETICAL BACKGROUND %%%
%%%%%%%%%%%%%%%%%%%%%%%%%%%%%%
\section{Theoretical background}
\label{sec:theory}

A system of free electrons and phonons can be described by the non-interacting Hamiltonian:
\begin{equation}
\label{eq:free-hamiltonian}
\hat{H}_0 
= \sum_{\mu \mathbf{k}} \epsilon_{\mu \mathbf{k}} \hat{c}_{\mu \mathbf{k}}^\dagger \hat{c}_{\mu \mathbf{k}}
+ \sum_{\lambda \mathbf{q}} \omega_{\lambda \mathbf{q}} 
\left(\hat{b}_{\lambda \mathbf{q}}^\dagger \hat{b}_{\lambda \mathbf{q}} + \frac{1}{2}\right),
\end{equation}
where $\hat{c}_{\mu \mathbf{k}}$ and $\hat{c}^\dagger_{\mu \mathbf{k}}$ are the electron annihilation and creation operators for state $\mu$ with momentum $\mathbf{k}$ and energy $\epsilon_{\mu \mathbf{k}}$; and where $\hat{b}_{\lambda \mathbf{q}}$ amd $\hat{b}^\dagger_{\lambda \mathbf{q}}$ are the phonon annihilation and creation operators for state $\lambda$ with momentum $\mathbf{q}$ and frequency $\omega_{\lambda\mathbf{q}}$.

In this section, we use the equilibrium Matsubara Green's function formalism to study the role of electron-phonon and phonon-phonon interactions on top of the non-interacting Hamiltonian in Eq.\,(\ref{eq:free-hamiltonian}). In Sec.\,\ref{sec:free-GF} we introduce the Green's functions for free electrons and free phonons. In Sec.\,\ref{sec:el-ph} we explore the extension of Eq.\,(\ref{eq:free-hamiltonian}) with the incorporation of linear electron-phonon coupling; and in Sec.\,\ref{sec:ph-ph} we explore the extension of Eq.\,(\ref{eq:free-hamiltonian}) with the incorporation of anharmonic phonon-phonon interactions. We bring these results together in Sec.\,\ref{sec:anh-ph-el}, where we explore anharmonic dephasing in the electron-phonon interaction.

%In this work, we adopt the imaginary-time formalism and use Matsubara Green’s functions throughout. This approach provides a convenient and powerful framework for studying systems in thermal equilibrium. Unlike real-time Green’s functions, which often suffer from oscillatory behavior, Matsubara Green’s functions are defined on the imaginary time axis and can be analytically continued to imaginary frequencies. This makes them particularly well-suited for capturing equilibrium properties and facilitates computations in the frequency domain.

% Green's functions for free electrons and phonons
\subsection{Green's Functions for free Electrons and Phonons}
\label{sec:free-GF}

The Matsubara Green’s functions for free electrons and phonons in thermal equilibrium are respectively defined as\,\cite{Mahan2000}: 
\begin{equation} 
G_{\mu \mathbf{k}}^0(\tau) = - i \langle 
\hat{T}_{\tau} 
\hat{c}_{\mu \mathbf{k}}(\tau) 
\hat{c}^\dagger_{\mu \mathbf{k}}(0) \rangle, 
\end{equation} 
\begin{equation} 
D_{\lambda \mathbf{q}}^0(\tau) = - i \langle 
\hat{T}_{\tau} 
\hat{A}_{\lambda \mathbf{q}}(\tau) 
\hat{A}^\dagger_{\lambda \mathbf{q}}(0) \rangle, 
\end{equation} 
where $\tau = it$ is the imaginary time, $\hat{T}_\tau$ is the imaginary time ordering operator, and $\hat{A}_{\lambda \mathbf{q}}(\tau) = \hat{b}_{\lambda \mathbf{q}}e^{-\omega_{\lambda \textbf{q}}\tau} + \hat{b}^\dagger_{\lambda, -\mathbf{q}}e^{\omega_{\lambda, -\textbf{q}}\tau}$ is the phonon field operator for mode $\lambda$ and momentum $\mathbf{q}$.
% The independent Matsubara Green’s functions of electron and phonon for a system at thermal equilibrium are defined as\,\cite{Mahan2000}:
% \begin{equation}
% G_{\mu \mathbf{k}}^0(\tau) = - i \langle \hat{T}_{\tau} 
% {\hat{c}}_{\mu \mathbf{k}}(\tau) {\hat{c}}^\dagger_{\mu \mathbf{k}}(0) \rangle,
% \end{equation}
% \begin{equation}
% D_{\lambda \mathbf{q}}^0(\tau) = - i \langle \hat{T}_{\tau} 
% {\hat{A}}_{\lambda \mathbf{q}}(\tau)  {\hat{A}}^\dagger_{\lambda \mathbf{q}}(0) \rangle,
% \end{equation}
% where $\tau=it$ is the imaginary time, $\hat{c}_{\mu \mathbf{k}}$ and $\hat{c}^\dagger_{\mu \mathbf{k}}$ are electron annihilation and creation operators for state $\mu$ with momentum $\mathbf{k}$, and $\hat{A}_{\lambda \mathbf{q}}(\tau) = \hat{b}_{\lambda \mathbf{q}}e^{-\omega_{\lambda \textbf{q}}\tau} + \hat{b}^\dagger_{\lambda-\mathbf{q}}e^{\omega_{\lambda -\textbf{q}}\tau}$ is the phonon field operator for mode $ \lambda $ with momentum $ \mathbf{q} $. 
% Here, $ \hat{T}_\tau $ denotes time ordering along the imaginary-time contour. Adopting the imaginary-time formalism and Matsubara Green’s functions provides a convenient framework for equilibrium systems.
% Imaginary-time Green’s functions efficiently capture equilibrium dynamics in the frequency domain, avoiding the oscillatory behavior of real-time Green’s functions.

In frequency space, the free particle electron Green's function is:
\begin{equation}
G_{\mu \mathbf{k}}^0(i\omega_n) = \frac{1}{i\omega_n - \epsilon_{\mu \mathbf{k}}},
\label{elgf}
\end{equation}
where $i\omega_n$ is the fermionic Matsubara frequency with $i\omega_n = (2n+1)\pi/\beta$ for $n\in\mathbb{Z}$, corresponding to antisymmetric states and Fermi-Dirac statistics. The corresponding free particle phonon Green's function is:
\begin{equation}
D_{\lambda \mathbf{q}}^0(i\omega_m) = \frac{-2\omega_{\lambda \mathbf{q}}}{\ \omega_{m}^2 + \omega_{{\mathbf{q}} \lambda}^2\ },
\label{phgf}
\end{equation}
where $i\omega_m$ is the bosonic Matsubara frequency with $i\omega_m = 2m\pi/\beta$ for $m \in \mathbb{Z}$, corresponding to symmetric states and Bose-Einstein statistics.

% Electron-phonon interactions
\subsection{Electron-Phonon Interactions in the Green's Function Formalism}
\label{sec:el-ph}

%The interactions of electrons in solids are exceptionally complex due to strong electron correlations and the diverse mechanisms of interaction, such as spin, charge, lattice, and orbital effects.
%To pave the way for the subsequent consideration of electron-phonon interactions, we adopt the ground-state mean-field approximation, i.e., DFT in our discussion, as the single-particle starting point.
%The Kohn-Sham auxiliary system provides a complete set of single-particle electron states with a high level of descriptive accuracy.
The standard electron-phonon Hamiltonian reads:
%This facilitates studying electron-phonon interactions in frequency space, where the Hamiltonian in the Heisenberg picture is given by  
\begin{equation}
\label{eq:elph}
    \hat{H}_{\mathrm{ep}} = \hat{H}_0 + \hat{V}_{\mathrm{ep}},
\end{equation}
where the electron-phonon interaction is given by:
\begin{equation}
\begin{split}
\hat{V}_{\mathrm{ep}} = \sum_{j=1}^{\infty} \frac{1}{\sqrt{N^{j}}} 
\sum_{\mu_2\mu_1 \mathbf{k}} 
\sum_{\lambda_1 \mathbf{q}_1, ..., \lambda_j \mathbf{q}_j} 
g_{\mu_2 \mu_1 \lambda_1 ... \lambda_j}^{\mathbf{k} \mathbf{q}_1 ... \mathbf{q}_j} \\
\hat{c}^\dagger_{\mu_2\mathbf{k}+\mathbf{q}_1+...+\mathbf{q}_j} \hat{c}_{\mu_1 \mathbf{k}} \hat{A}_{\lambda_1 \mathbf{q}_1} ... \hat{A}_{\lambda_j \mathbf{q}_j},
\end{split}
\end{equation}
% where $g_{\mu_2 \mu_1 \lambda}^{\mathbf{k} \mathbf{q}}$ are the electron-phonon coupling coefficients and $N$ is the number of unit cells in the system. 
where $g_{\mu_2 \mu_1 \lambda_1 ... \lambda_j}^{\mathbf{k} \mathbf{q}_1 ... \mathbf{q}_j}$ are the electron-phonon coupling coefficients with $j$ phonon lines corresponding to the $j$th order coupling coefficient, and $N$ is the number of unit cells in the system. 

\begin{figure}
    \centering
    \includegraphics[width=\linewidth]{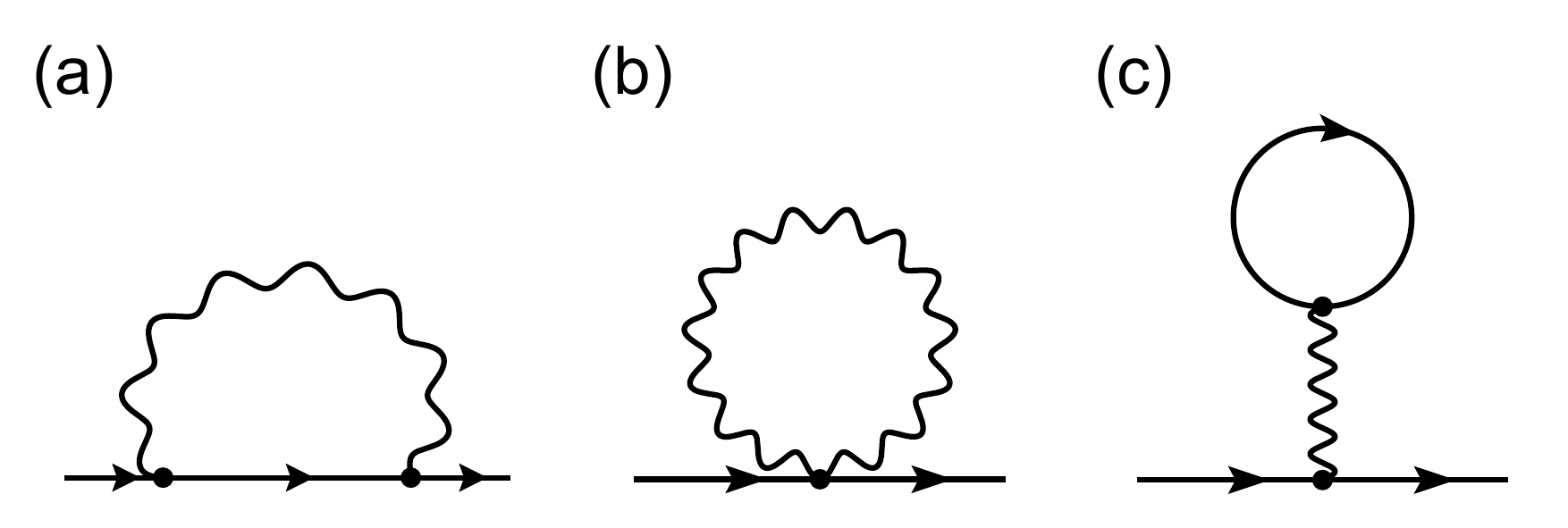}
    \caption{Electron-phonon coupling Feynman diagrams within second-order perturbation theory. 
        (a) Fan-Migdal diagram with two first-order electron-phonon coupling vertices;
        (b) Debye-Waller diagram with one second-order electron-phonon coupling vertex;
        (c) tadpole diagram with two first-order electron-phonon coupling vertices.
    }
    \label{fig:epi_diagrams}
\end{figure}

%In the Hamiltonian in Eq.\,(\ref{eq:elph}), the interaction $\hat{V}$ is given by the third term representing electron-phonon coupling. The exact electron Green's function for the electron-phonon system can be written as an expansion in terms of the interaction:
%\begin{equation}
%\begin{split}
%G_{\mu \mathbf{k}}(\tau) = - \sum_{n=1}^\infty 
%&\left(-\frac{1}{\hbar}\right)^n\int_{0}^{\beta\hbar} 
%d\tau_1 ... \int_{0}^{\beta\hbar}d\tau_n \\
%&\langle T_\tau \hat{c}_{\mu {\mathbf{k}}}(\tau) \hat{c}^\dagger_{\mu {\mathbf{k}}}(0) \hat{V}(\tau_1) ... \hat{V}(\tau_n) \rangle_0.
%\end{split}
%\label{eq:el_gf}
%\end{equation}
%In practice, this interacting Green's function can be evaluated 
%using a perturbative expansion and exploiting Wick's theorem, which enables the decomposition of Eq.\,(\ref{eq:el_gf}) into an infinite sum of coupling vertices and independent propagators. This expansion is managed 
% where $\hat{V}$ corresponds to the interaction part of the Hamiltonian.
% The computation of the interacting Green's function relies on a perturbative expansion based on Wick's theorem to decompose Eq.\,\ref{eq:el_gf} into an infinite number of linear combinations of the product of several coupling vertices and independent particle propagators.
% One way to deal with this expansion is to define the irreducible self-energy $\Sigma_{\mu \mathbf{k}}(\tau)$ as
The interacting electron Green's function is given by the Dyson equation:
\begin{equation}
G_{\mu \mathbf{k}}(i\omega_n) =  
G_{\mu \mathbf{k}}^0(i\omega_n) + 
G_{\mu \mathbf{k}}^0(i\omega_n) 
\Sigma_{\mu \mathbf{k}}(i\omega_n)
G_{\mu \mathbf{k}}(i\omega_n),
\end{equation}
where $\Sigma_{\mu \mathbf{k}}(i\omega_n)$ is the irreducible self-energy arising from the electron-phonon interaction. 

In practice, $\Sigma_{\mu \mathbf{k}}(\tau)$ can be approximated by using a low-order perturbative expansion. The corresponding terms including perturbations up to second order in the interaction are depicted as Feynman diagrams in Fig.\,\ref{fig:epi_diagrams}.
The diagram in Fig.\,\ref{fig:epi_diagrams}(a) is the Fan-Migdal self-energy, a frequency-dependent dynamical contribution that affects both the energy levels (real part) and lifetimes (imaginary part) of electronic states. This term has been extensively studied and shown to be essential for understanding superconductivity\,\cite{Margine2013}, charge transport\,\cite{Chang2022}, and other key material properties\,\cite{Antonius2014,Monserrat2016,Zhou2018}.

Figure~\ref{fig:epi_diagrams}(b) is the Debye-Waller self-energy. This term, arising from the second-order electron-phonon coupling coefficients, is independent of frequency and purely real, resulting in a rigid shift in the electronic bands. Physically, it accounts for the average potential felt by electrons due to thermal lattice vibrations, and it depends only on the phonon population and electronic density of states, rather than the dynamical response of the system.

% Lastly, the tadpole diagram in Fig.\,\ref{fig:epi_diagrams}(c) is usually overlooked because it is zero in centrosymmetric systems. However, this term becomes important in the study of polarons, where it corresponds to polaronic lattice distortions\,\cite{Lafuente2022PRB}. Polarons distort the equilibrium ionic positions, breaking the original symmetry and leading to a non-zero tadpole contribution.

Lastly, the tadpole diagram in Fig.\,\ref{fig:epi_diagrams}(c) is zero in our formulation.
Within the Born–Oppenheimer approximation, the electronic Hamiltonian is defined at the DFT equilibrium geometry, where the total derivative of the Born–Oppenheimer energy with respect to the ionic coordinates vanishes.
At this point, the electron–phonon tadpole contribution is exactly canceled by the corresponding nucleus–nucleus term, as shown in Ref.\,\cite{Marini2015}.
But this term is still vital in the discussion of polaronic lattice distortions\,\cite{Lafuente2022PRB}, phonon's renormalization of electron band structures, and many out-of-equilibrium phenomena\,\cite{Stefanucci2023}.

The combination of the Fan-Migdal, Debye-Waller, and tadpole diagrams constitutes the complete set of second-order perturbative corrections arising from electron-phonon interactions. This level of theory is considered sufficient for the description of a vast array of electron-phonon driven material properties. Qualitatively, this can be justified by the Migdal approximation\,\cite{Migdal1958}, which asserts that vertex corrections can be neglected in systems where the electron mass is much smaller than the ion mass. %, leading to a small adiabatic parameter $\eta = \frac{\omega_{ph}}{E_f}$.
% However, in recent years, a growing body of research, particularly in superconducting community, has begun to challenge the validity of the Migdal approximation, especially in systems with strong electron-phonon coupling\,\cite{Schrodi2020}, low carrier densities\,\cite{Phan2022}, or near quantum critical points\,\cite{Jarlborg2016}, where vertex corrections and higher-order processes can no longer be ignored.

Next, we derive the formula for the Fan-Migdal self-energy in Fig.\,\ref{fig:epi_diagrams}(a)
%, which corresponds to the lowest-order term in the electron-phonon interaction. 
While this derivation can be found in standard textbooks\,\cite{Mahan2000}, it will serve as a starting point for our subsequent derivation of anharmonic dephasing in electron-phonon coupling. We first recall the Feynman diagram rules for the electron-phonon coupling problem, following Sec.\,3.4 in Ref.\,\cite{Mahan2000}:
%In our convention, unprimed indices label external lines, while primed indices represent internal degrees of freedom to be summed over in the perturbative expansion:
\begin{itemize}
\item Draw all the topologically distinct connected diagrams up to order $n$, where the order refers to the number of phonon lines.
In each diagram, an incoming free-electron line labeled $\mu_\mathrm{in}\mathbf{k}_\mathrm{in}$ enters from the left of the page and an outgoing free-electron line labeled $\mu_\mathrm{out} \mathbf{k}_\mathrm{out}$ leaves at the right of the page. Arrows indicate the direction of propagation of the particles.
%\textcolor{red}{to do: Add topological distinct diagrams ...; rephrase the formulas according to Mahan's book}
\item Associate each internal electron line with a non-interacting electron Green's function $G_{\mu \mathbf{k}}^0(i\omega_n)$, as given in Eq.\,(\ref{elgf}).
\item Associate each internal phonon line with a non-interacting phonon Green's function $D_{\lambda \mathbf{q}}^0(i\omega_n)$, as given in Eq.\,(\ref{phgf}).
\item Associate each vertex with an electron-phonon coupling coefficient $g_{\mu_2 \mu_1 \lambda}^{\mathbf{k} \mathbf{q}}$.
\item Conserve momentum and complex frequency at each vertex. The oddness and evenness of fermion frequencies and boson frequencies are maintained in the energy conservation.
\item Sum over all the internal degrees of freedom: momentum and frequency. %Internal variables are all those except the $(\mu\mathbf{k},i\omega_n)$.
\item Multiply the result by the factor:
\begin{equation}
    \left(-\frac{1}{\beta N}\right)^n (-2)^F,
\end{equation}
where $F$ is the number of closed fermion loops and $n$ is the order of the diagram.
\end{itemize}

Following these rules, the 
%diagonal terms of the 
Fan-Midgal self-energy depicted in Fig.\,\ref{fig:epi_diagrams}(a) corresponds to order $n=1$.
To simplify notation, we drop the %earlier convention of priming internal degrees of freedom and 
labeling of external lines as ``in'' and ``out'', and the Fan-Midgal self-energy can be expressed as:
\begin{equation}
\begin{split}
\Sigma^\text{FM}_{\mu\mathbf{k}}(i\omega_m) = 
-\frac{1}{\beta N} 
&
\sum_{\mu_1}
\sum_{\lambda \mathbf{q}} 
\sum_{i\omega_{n}} 
g_{\mu_1\mu\lambda}^{\mathbf{kq}}
D^{0}_{\lambda\mathbf{q}}(i\omega_n) \\
&G^{0}_{\mu_1\mathbf{k}+\mathbf{q}}(i\omega_m + i\omega_n)
g_{\mu_1\mu\lambda}^{\mathbf{k} \mathbf{q}*}.
\end{split}
\end{equation}
Substituting the free-particle Green's functions into the self-energy, it can be re-written as:
\begin{equation}
\Sigma^\text{FM}_{\mu\mathbf{k}}(i\omega_m) = \frac{1}{N}
\sum_{\mu_1}
\sum_{\lambda \mathbf{q}}
|g_{\mu_1\mu\lambda}^{\mathbf{k} \mathbf{q}}|^2
\left(-\frac{1}{\beta}\right)
\sum_{i\omega_n}
f(i\omega_n),
\end{equation}
where $f(z)$ is given by:
\begin{equation}
f(z) = 
\frac{2\omega_{\lambda \mathbf{q}}}{z^2 - \omega_{\lambda \mathbf{q}}^2}
\frac{1}{z + i\omega_{m} - \varepsilon_{\mu_2\mathbf{k}+\mathbf{q}}}.
\end{equation}
To evaluate this expression, we use the Matsubara frequency summation technique and the application of the Cauchy residue theorem:
\begin{equation}
\begin{split}
0 
&= \lim_{|z| \to \infty} 
\oint \frac{dz}{2 \pi i} f(z) n_\mathrm{B}(z)  \\
&= \sum_{z}^{f \cdot n_\mathrm{B}} 
\text{Res} \{f(z) n_\mathrm{B}(z) \} \\
&= 
\sum_{i\omega_n}
f(i\omega_n) 
\frac{1}{\beta} + 
\sum_{z}^{f}
\text{Res} \{ f(z) n_\mathrm{B}(z) \},
\end{split}
\end{equation}
where $n_\mathrm{B}$ is the bosonic weighting function, defined as the analytic continuation of the Bose–Einstein distribution given as:
\begin{equation}
    n_\mathrm{B}(z) = \frac{1}{e^{\beta z} - 1},
\end{equation}
and the poles of $f(z)$ are at $z_1 = \omega_{\lambda \mathbf{q}}$, $z_2 = -\omega_{\lambda \mathbf{q}}$, and $z_3 = -i\omega_{m} + \varepsilon_{\mu_2\mathbf{k}+\mathbf{q}}$, whose corresponding residues are:
\begin{equation}
\mathrm{Res}\{f(z),z_1\} = 
\frac{1}{\omega_{\lambda \mathbf{q}} + i\omega_m - \varepsilon_{\mu_2k+\mathbf{q}}},
\end{equation}
\begin{equation}
\mathrm{Res}\{f(z),z_2\} = 
-
\frac{1}{-\omega_{\lambda \mathbf{q}} + i\omega_m - \varepsilon_{\mu_2k+\mathbf{q}}},
\end{equation}
\begin{equation}
\begin{split}
\mathrm{Res}\{f(z),z_3\} = 
&\frac{1}{-\omega_{\lambda \mathbf{q}} + i\omega_m - \varepsilon_{\mu_2k+\mathbf{q}}}\\
&-\frac{1}{\omega_{\lambda \mathbf{q}} + i\omega_m - \varepsilon_{\mu_2k+\mathbf{q}}}.
\end{split}
\end{equation}

Therefore, the self-energy term can be finally formulated as:
\begin{equation}
\label{eq:elph-self-energy}
\begin{split}
\Sigma^{\text{FM}}_{\mu\mathbf{k}}(\varepsilon_{\mu \mathbf{k}}) = \frac{1}{N} 
&\sum_{\mu_1}\sum_{\lambda\mathbf{q}}
|g_{\mu_1\mu\lambda}^\mathbf{kq}|^2 \\
\times & \Big[ 
\frac{\mathcal{N}_{\lambda \mathbf{q}} + f_{\mu_1 \mathbf{k}+\mathbf{q}}}
{\varepsilon_{\mu \mathbf{k}} + \omega_{\lambda \mathbf{q}} - \varepsilon_{\mu_1k+\mathbf{q}} + i\eta} \\
&+\frac{1 + \mathcal{N}_{\lambda \mathbf{q}} - f_{\mu_1 \mathbf{k}+\mathbf{q}}}
{\varepsilon_{\mu \mathbf{k}} - \omega_{\lambda \mathbf{q}} - \varepsilon_{\mu_1\mathbf{k}+\mathbf{q}} + i\eta}
\Big],
\end{split}
\end{equation}
where we use analytic continuation $i\omega_{m} \to \varepsilon + i\eta$ within the off-shell approximation.

Starting from the Fan-Migdal self-energy in Eq.\,(\ref{eq:elph-self-energy}), we can construct the associated scattering rate $\Gamma^\text{FM}_{\mu \mathbf{k}}$, %which is the inverse of the lifetime, 
given by the imaginary part of the self-energy.
Working within the self-energy relaxation time approximation, and using the identity $\frac{1}{x+i\eta} = P \frac{1}{x} - i\pi \delta(x)$, we obtain:
\begin{equation}
\begin{split}
\Gamma^\text{FM}_{\mu \mathbf{k}} &= 
\frac{2\pi}{N}
\sum_{\mu_1} \sum_{\lambda \mathbf{q}}
|g_{\mu_1\mu\lambda}^\mathbf{kq}|^2 \\
\times & \Big[ (\mathcal{N}_{\lambda \mathbf{q}} + f_{\mu_1 \mathbf{k}+\mathbf{q}}) 
\delta(\varepsilon_{\mu \mathbf{k}} +  \omega_{\lambda \mathbf{q}} - \varepsilon_{\mu_1 \mathbf{k}+\mathbf{q}}) \\
&+(1 + \mathcal{N}_{\lambda \mathbf{q}} - f_{\mu_1 \mathbf{k}+\mathbf{q}}) 
\delta(\varepsilon_{\mu \mathbf{k}} -  \omega_{\lambda \mathbf{q}} - \varepsilon_{\mu_1 \mathbf{k}+\mathbf{q}})
\Big].
\end{split}
\end{equation}
The Fan-Migdal scattering rate $\Gamma^\text{FM}_{\mu \mathbf{k}}$ captures how likely an electron in band $\mu$ and momentum $\mathbf{k}$ is to scatter by emitting or absorbing phonons, leading to a finite lifetime for that quasiparticle state.

The derivation in this section illustrates the standard strategy to study many-body interacting systems at equilibrium. We can divide it into three stages: (i) building a Hamiltonian that includes the target interaction; (ii) identifying the relevant self-energy terms in the perturbative expansion; and (iii) deriving the associated scattering rates. We will again use this strategy to describe anharmonic phonon-phonon interactions in Sec.\,\ref{sec:ph-ph}, and then to describe anharmonic dephasing in electron-phonon interactions in Sec.\,\ref{sec:anh-ph-el}.

% Anharmonic phonon-phonon interactions
\subsection{Phonon-Phonon Interactions in the Green's Function Formalism} 
\label{sec:ph-ph}

Ionic vibrations are typically described within the harmonic approximation, which corresponds to a second order expansion of the Born-Oppenheimer energy surface about equilibrium. The harmonic approximation directly leads to non-interacting phonons as in Eq.\,(\ref{eq:free-hamiltonian}). The inclusion of anharmonic phonon-phonon interactions requires the inclusion of higher-order terms in the expansion of the Born-Oppenheimer energy surface, leading to the following anharmonic phonon-phonon Hamiltonian\,\cite{Born1996,Leibfried1961}:
\begin{equation}
\label{eq:phph}
    \hat{H}_{\mathrm{pp}}=\hat{H}_0 + \hat{V}_{\mathrm{pp}},
\end{equation}
where the phonon-phonon interaction is given by:
\begin{equation}
%\begin{split}
\hat{V}_{\textrm{pp}} = \sum_{j=3}^{\infty} %&
\frac{N^{1-j/2}}{j!}\!\!\!\sum_{\lambda_1{q}_1,...,\lambda_j{q}_j}\!\!\!
\phi^{\mathbf{q}_1\mathbf{q}_2...\mathbf{q}_j}_{\lambda_1\lambda_2...\lambda_j} \\
%&
\hat{A}_{\lambda_1 \mathbf{q}_1} \hat{A}_{\lambda_2 \mathbf{q}_2}... \hat{A}_{\lambda_j \mathbf{q}_j},
%\end{split}
\end{equation}
where $j$ is the number of the phonons participating in the scattering. The associated $j$th order phonon-phonon coupling coefficients are given by:
\begin{equation}
%\begin{split}
\phi^{\mathbf{q}_1\mathbf{q}_2...\mathbf{q}_j}_{\lambda_1\lambda_2...\lambda_j}
=\!\!\sum_{\substack{\kappa_1...\kappa_j\\ \alpha_1...\alpha_j}}\!\!
\Phi_{\kappa_1...\kappa_j}^{\alpha_1...\alpha_j}(\mathbf{q}_1,...,\mathbf{q}_j)\!\prod_{i=1}^j\!\sqrt{\frac{1}{2M_{\kappa_i}\omega_{\lambda_i\mathbf{q}_i}}}\epsilon_{\lambda_i\kappa_i}^{\alpha_i\mathbf{q}_i},
%\\
%&\times\sqrt{\frac{\hbar}{2M_{\kappa_1}\omega_{\lambda_1\mathbf{q}_1}}}...\sqrt{\frac{\hbar}{2M_{\kappa_j}\omega_{\lambda_j\mathbf{q}_j}}}\times \epsilon_{\lambda_1\kappa_1}^{\alpha_1\mathbf{q}_1} ... \epsilon_{\lambda_j\kappa_j}^{\alpha_j\mathbf{q}_j},
\label{anharmonic coefficient}
%\end{split}
\end{equation}
where $\epsilon_{\lambda_i\kappa_i}^{\alpha_i\mathbf{q}_i}$ is a phonon eigenvector labelled by atom $\kappa_i$ of mass $M_{\kappa_i}$ and displacement direction $\alpha_i$; and $\Phi_{\kappa_1...\kappa_j}^{\alpha_1...\alpha_j}(\mathbf{q}_1,...,\mathbf{q}_j)$ is the Fourier transform of the $j$th order matrix of force constants. Momentum conservation is fullfilled modulo a reciprocal lattice vector $\mathbf{G}$ as $\mathbf{q}_1+\mathbf{q}_2+...+\mathbf{q}_j=\mathbf{G}$. % and $\mathbf{G}$.

\begin{figure*}
\centering
\includegraphics[width=\linewidth]{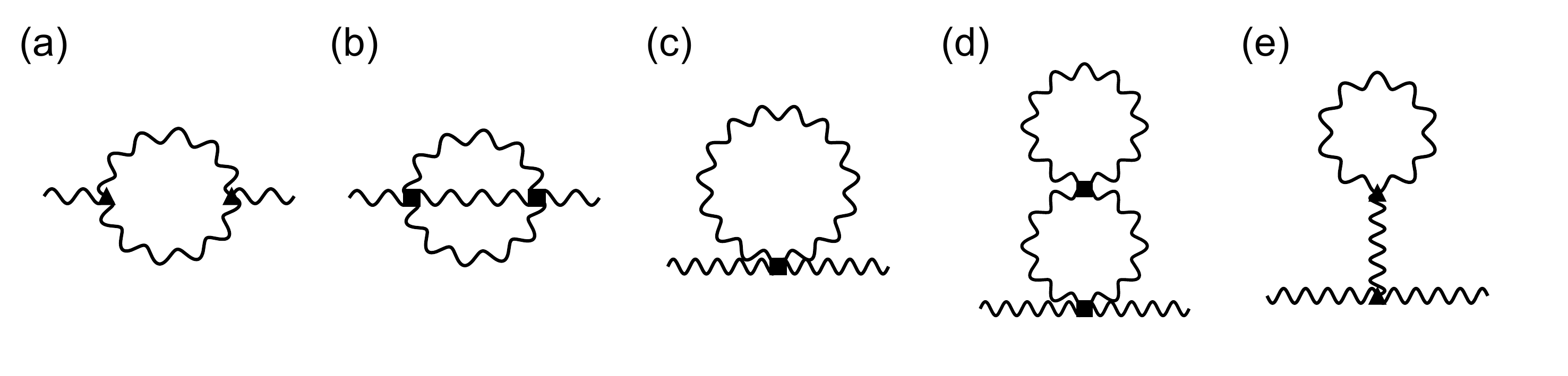}
\caption{
Feynman diagrams for phonon-phonon interactions.
(a) Second-order bubble diagram driven by three-phonon coupling; 
(b) second-order bubble diagram driven by four-phonon coupling;
(c) first-order loop diagram driven by four-phonon coupling;
(d) second-order loop diagram driven by four-phonon coupling;
(e) second-order tadpole diagram driven by three-phonon coupling. 
}
\label{fig:ppi_diagrams}
\end{figure*}

Following the same strategy as in Sec.\,\ref{sec:el-ph}, we can express the interacting phonon Matsubara Green's function with the Dyson equation as:
\begin{equation}
D_{\lambda\mathbf{q}}(i\omega_m) = D^0_{\lambda\mathbf{q}}(i\omega_m) + 
D^0_{\lambda\mathbf{q}}(i\omega_m) \Pi_{\lambda\mathbf{q}}(i\omega_m) D_{\lambda\mathbf{q}}(i\omega_m),
\end{equation}
where $\Pi_{\lambda\mathbf{q}}(\tau)$ is the irreducible self-energy arising from the phonon-phonon interactions. 
Each irreducible self-energy term can be represented by a Feynman diagram and we show several leading-order diagrams in Fig.\,\ref{fig:ppi_diagrams}.
Each of these Feynman diagrams represents a unique aspect of the physical processes underlying the anharmonic crystal.

The bubble diagram mediated by the three-phonon coupling in Fig.\,\ref{fig:ppi_diagrams}(a) captures the interaction between real, thermally excited phonons. It represents a resonant scattering process in which a phonon interacts with another phonon present in the thermal bath, leading to an intermediate two-phonon state before recombining. 
This process gives rise to both a frequency renormalization and a finite phonon lifetime, corresponding to the real and imaginary parts of the phonon self-energy, respectively.
Therefore, it accounts for anharmonic dephasing of phonons, playing a crucial role in thermal transport and other finite-temperature lattice properties.

Figure~\ref{fig:ppi_diagrams}(b) shows the corresponding four-phonon bubble diagram. Like its three-phonon counterpart in Fig.\,\ref{fig:ppi_diagrams}(a), it contributes to both the real and imaginary parts of the phonon self-energy. Notably, this diagram has been shown to significantly affect thermal transport in materials with large phonon band gaps\,\cite{Feng2016,Fei2018,Ravichandran2020}. Therefore, an increasing number of studies are beginning to incorporate four-phonon scattering processes into first principles calculations to achieve more accurate predictions of thermal conductivity.

The loop diagram mediated by the four-phonon coupling in Fig.\,\ref{fig:ppi_diagrams}(c) contributes to the energy modification by capturing the anharmonic renormalization of phonon frequencies due to virtual scattering processes. 
Physically, this diagram describes a phonon that momentarily decays into a pair of virtual phonons, which then recombine to restore the original mode.
Such self-energy corrections arise from the quartic terms in the lattice potential energy and represent a uniquely quantum process arising from virtual phonon exchange, which has no counterpart in classical lattice dynamics.
This term accounts for the temperature dependence of phonon frequencies, which can lead to either softening or hardening. 
A similar higher-order interaction is depicted in Fig.\,\ref{fig:ppi_diagrams}(d). 
These corrections are typically captured through variational approaches to avoid divergences and violations of causality that can arise from perturbative expansions\,\cite{Tadano2015}.

% The tadpole diagram in Fig.\,\ref{fig:ppi_diagrams}(e) is usually overlooked. It represents a static change to the lattice equilibrium position rather than a real phonon scattering process. 
% It arises from the contraction of third-order force constants with internal phonon fluctuations and reflects how anharmonicity, particularly non-centrosymmetric terms, can shift the atomic equilibrium positions. Specifically, if the intermediate vertical phonon line corresponds to an optical phonon it accounts for the relaxation of internal coordinates; and if the intermediate line is an acoustic phonon line, then it corresponds to the relaxation of cell parameters\,\cite{Paulatto2015}. Since it involves no energy or momentum exchange, its contribution does not affect phonon lifetimes. 
% In many cases, especially for centrosymmetric systems, this term vanishes or can be absorbed during structural relaxation, which is why the tadpole diagram is often neglected in practical calculations. 
% However, in anharmonic phonon systems, temperature-dependent corrections shift the vibrational equilibrium, and therefore this self-energy term needs to be properly accounted for.

The tadpole diagram in Fig.\,\ref{fig:ppi_diagrams}(e) represents a static renormalization of the lattice equilibrium positions rather than a genuine phonon-scattering process. 
It originates from the contraction of third-order force constants with the internal phonon fluctuations and describes how anharmonicity shifts the vibrational equilibrium geometry. 
Depending on the phonon branch associated with the internal line, this term accounts for the relaxation of internal coordinates (optical modes) or of the cell parameters (acoustic modes),\cite{Paulatto2015}.
In a perturbative expansion around the equilibrium harmonic structure, the phonon–phonon tadpole diagram vanishes exactly. 
This cancellation follows from the permutation symmetry of the third-order force constants together with momentum conservation at the interaction vertex. 
Consequently, the corresponding contribution is absorbed into the structural relaxation that defines the harmonic reference geometry and does not affect phonon lifetimes.
However, when the system is displaced away from its equilibrium geometry—for example in strongly anharmonic materials or in temperature-dependent renormalizations—the vibrational equilibrium shifts and tadpole contributions must be properly taken into account.

% Following the same rules of using primed indices for internal variables, t
Next, we derive the formula for the three-phonon bubble diagram in Fig.\,\ref{fig:ppi_diagrams}(a),
%, which corresponds to the lowest-order term in the electron-phonon interaction. 
for which we first recall the associated Feynman diagram rules\,\cite{Maradudin1962}:
\begin{itemize}
    \item Draw all topologically distinct connected $n$th-order diagrams in which a free phonon line labeled by $\lambda_\mathrm{in}\mathbf{q}_\mathrm{in}$ enters from the left of the page and a free phonon line labeled $\lambda_\mathrm{out} \mathbf{q}_\mathrm{out}$ leaves at the right of the page. %Although it is not necessary to direct the phonon lines, it is convenient to direct all lines right.
    \item With each internal phonon line labeled $\lambda \mathbf{q}$, associate a factor 
    \begin{equation}
    D_{\lambda \mathbf{q}}(i\omega_l) = \frac{2\omega_{\lambda \mathbf{q}}}{\beta} \frac{1}{\omega_l^2 + \omega^2_{\lambda \mathbf{q}}}, \quad i\omega_l = \frac{2\pi l}{\beta}.
    \end{equation}
    \item At each vertex, conserve the $\mathbf{q}$ vectors according to the rule that the sum of the wave vectors for lines leaving the vertex equals the sum of the wave vectors for lines entering the vertex, modulo a reciprocal lattice vector.
    \item At each vertex, conserve the frequencies $\omega_l$ according to the rule that the sum over the frequencies $\omega_l$ leaving the vertex equals the sum over frequencies $\omega_l$ entering the vertex.
    \item At each vertex, insert the appropriate matrix element.
    \item Insert a factor $\left[(-1)^n / n! \right] \beta^n$, where $\beta^n$ comes from the integrations over the $n$ $\beta$-variables.
    \item Insert a combinatorial factor which gives the number of pairing schemes to which the diagram corresponds. This factor is the product of the number of topologically equivalent diagrams that can be drawn for a fixed arrangement of the vertices, the number of different labelings at each vertex for each pairing scheme, and the number of ways of permuting the order of the phonon vertices.
    \item Finally, sum over the independent indices.
\end{itemize}

% Based on these rules, we can directly write the formulae associated with the leading order Feynman diagrams shown in Fig.\,\ref{fig:ppi_diagrams}. Each of these Feynman diagrams represents a unique aspect of the physical processes underlying the anharmonic crystal.

%, the correction in Fig.\,\ref{fig:ppi_diagrams}(c) usually won't be added isolately. 
%On the other hand, we incorporate it through variational approaches to avoid divergences and violations of causality that can arise from naive perturbative inclusion.
%For example, Fig.\,\ref{fig:ppi_diagrams}(d) is a demonstration of some implementation of self-consistent phonon theory\,\cite{Tadano2015}. 
%Essentially, it incorporates the loop diagram into the intermediate loop-phonon line, allowing additional loops to be consistently added to the bare loop in a self-consistent manner.

Based on these rules, we can directly write the formulae associated with the leading order Feynman diagrams shown in Fig.\,\ref{fig:ppi_diagrams}.
The lowest-order phonon-phonon diagram giving rise to a finite phonon lifetime is that in Fig.\,\ref{fig:ppi_diagrams}(a), which is the focus of our work. Mirroring the discussion of the electron-phonon interaction in Sec.\,\ref{sec:el-ph}, we give a brief derivation for the three-phonon bubble diagram in Fig.\,\ref{fig:ppi_diagrams}(a).
Directly using the Feynman rules, we can write the bubble self-energy as:
\begin{equation}
\begin{split}
    \Pi^{\mathrm{B}}_{\lambda\mathbf{q}}&(i\omega_l)
    =
    -\frac{1}{2N} \frac{1}{\beta} 
    \sum_{\substack{\lambda_1\mathbf{q}_1\\\lambda_2\mathbf{q}_2}} \sum_{i\omega_m}
    \phi_{\lambda\lambda_1\lambda_2}^{-\mathbf{q}\mathbf{q}_1\mathbf{q}_2}
    \phi_{\lambda\lambda_1\lambda_2}^{\mathbf{q},-\mathbf{q}_1,-\mathbf{q}_2} \\
    &\times 
    D^0_{\lambda_1\mathbf{q}_1}(i\omega_m)
    D^0_{\lambda_2\mathbf{q}_2}(i\omega_l-i\omega_m)
    \Delta_{\mathbf{q}-\mathbf{q}_1-\mathbf{q}_2},
\end{split}
\end{equation}
where $\Delta_{\mathbf{q}-\mathbf{q}_1-\mathbf{q}_2}$ imposes the condition $\mathbf{q}-\mathbf{q}_1-\mathbf{q}_2=\mathbf{G}$ on the phonon momenta, where $\mathbf{G}$ is a reciprocal lattice vector.
Substituting the free phonon Green's functions into this equation, we obtain:
\begin{equation}
    \Pi^{\mathrm{B}}_{\lambda\mathbf{q}}(i\omega_l) = 
    \frac{-1}{2N} 
    \sum_{\substack{\lambda_1\mathbf{q}_1\\\lambda_2\mathbf{q}_2}} 
    |\phi_{\lambda\lambda_1\lambda_2}^{-\mathbf{q}\mathbf{q}_1\mathbf{q}_2}|^2 
    \Delta_{\mathbf{q}-\mathbf{q}_1-\mathbf{q}_2}
    \frac{1}{\beta} \sum_{i\omega_m} f(i\omega_m),
\end{equation}
where $f(z)$ is given by:
\begin{equation}
    f(z) = 
    \frac{2\omega_{\lambda_1\mathbf{q}_1}}{z^2 - \omega_{\lambda_1\mathbf{q}_1}^2}
    \frac{2\omega_{\lambda_2\mathbf{q}_2}}{(z-i\omega_n)^2-\omega_{\lambda_2\mathbf{q}_2}^2}.
\end{equation}
We apply the Matsubara summation technique on the poles $z_1=\omega_{\lambda_1\mathbf{q}_1}$, $z_2=-\omega_{\lambda_1\mathbf{q}_1}$, $z_3=i\omega_n+\omega_{\lambda_2\mathbf{q}_2}$, and $z_4=i\omega_n-\omega_{\lambda_2\mathbf{q}_2}$; and obtain the self-energy as:
\begin{equation}
\begin{split}
\Pi^\mathrm{B}_{\lambda\mathbf{q}}(i\omega_l) = \frac{-1}{2N}
\sum_{\substack{\lambda_1\mathbf{q}_1\\\lambda_2\mathbf{q}_2}} 
|&\phi_{\lambda\lambda_1\lambda_2}^{-\mathbf{q}\mathbf{q}_1\mathbf{q}_2}|^2 
\Delta_{\mathbf{q}-\mathbf{q}_1-\mathbf{q}_2} \\
\left\{-
\frac{\mathcal{N}_{\lambda_1\mathbf{q}_1}+\mathcal{N}_{\lambda_2\mathbf{q}_2}+1}
{i\omega_{l}-\omega_{\lambda_1\mathbf{q}_1}-\omega_{\lambda_2\mathbf{q}_2}}
\right.
&+
\frac{\mathcal{N}_{\lambda_1\mathbf{q}_1}-\mathcal{N}_{\lambda_2\mathbf{q}_2}}
{i\omega_{l}-\omega_{\lambda_1\mathbf{q}_1}+\omega_{\lambda_2\mathbf{q}_2}} \\
-
\frac{\mathcal{N}_{\lambda_1\mathbf{q}_1}-\mathcal{N}_{\lambda_2\mathbf{q}_2}}
{i\omega_{l}+\omega_{\lambda_1\mathbf{q}_1}-\omega_{\lambda_2\mathbf{q}_2}}
&+
\left.\frac{\mathcal{N}_{\lambda_1\mathbf{q}_1}+\mathcal{N}_{\lambda_2\mathbf{q}_2}+1}
{i\omega_{l}+\omega_{\lambda_1\mathbf{q}_1}+\omega_{\lambda_2\mathbf{q}_2}}
\right\}.
\end{split}
\end{equation}
The associated scattering rate can then be written as:
\begin{equation}
\begin{split}
\Gamma^\mathrm{B}_{\lambda\mathbf{q}} = \frac{\pi}{N}
\sum_{\substack{\lambda_1\mathbf{q}_1\\\lambda_2\mathbf{q}_2}} 
|\phi_{\lambda\lambda_1\lambda_2}^{-\mathbf{q}\mathbf{q}_1\mathbf{q}_2}|^2 
&\Delta_{\mathbf{q}-\mathbf{q}_1-\mathbf{q}_2}\\
\times\left[
(\mathcal{N}_{\lambda_1\mathbf{q}_1} + \mathcal{N}_{\lambda_2\mathbf{q}_2} + 1)\right.
&\delta(\omega_{\lambda\mathbf{q}}-\omega_{\lambda_1\mathbf{q}_1}-\omega_{\lambda_2\mathbf{q}_2})\\
+2(\mathcal{N}_{\lambda_1\mathbf{q}_1}-\mathcal{N}_{\lambda_2\mathbf{q}_2})
&\left.\delta(\omega_{\lambda\mathbf{q}} - \omega_{\lambda_1\mathbf{q}_1} + \omega_{\lambda_2\mathbf{q}_2})
\right].
\end{split}
\end{equation}
The scattering rate $\Gamma^\mathrm{B}_{\lambda\mathbf{q}}$ associated with the bubble diagram captures the finite phonon lifetime caused by the phonon emission and absorption processes, reflecting the likelihood of phonons in this state being scattered by other phonons.

This derivation of the three-phonon bubble diagram provides another starting point for the study of phonon dephasing in electron-phonon coupling, discussed in Sec.\,\ref{sec:anh-ph-el} below.

\subsection{Electron-Phonon Interactions in Anharmonic Crystals}
\label{sec:elph-anh-overview}

The electron-phonon and phonon-phonon interactions are typically treated separately, as described above in Secs.\,\ref{sec:el-ph} and \ref{sec:ph-ph}, respectively. In this work, we combine both interactions and consider a Hamiltonian given by:
\begin{equation}
    \hat{H}_{\mathrm{eap}} = \hat{H}_0 + \hat{V}_{\mathrm{ep}} + \hat{V}_{\mathrm{pp}},
\end{equation}
where $\hat{V}_{\mathrm{ep}}$ and $\hat{V}_{\mathrm{pp}}$ are the interacting parts of the Hamiltonian and correspond to the electron-phonon coupling and to the phonon-phonon coupling given in Sec.\,\ref{sec:el-ph} and Sec.\,\ref{sec:ph-ph}, respectively.

The electron self-energy corrections arising from the inclusion of anharmonic phonon effects into the electron-phonon interaction can be divided into four categories: (i) finite phonon lifetimes induced by phonon-phonon scattering; (ii) anharmonic renormalization of phonon eigenvalues and eigenvectors; (iii) anharmonic corrections to the electron-phonon coupling vertex; and (iv) structural relaxation caused by anharmonic effects.

\begin{figure}[ht]
\centering
\includegraphics[width=\linewidth]{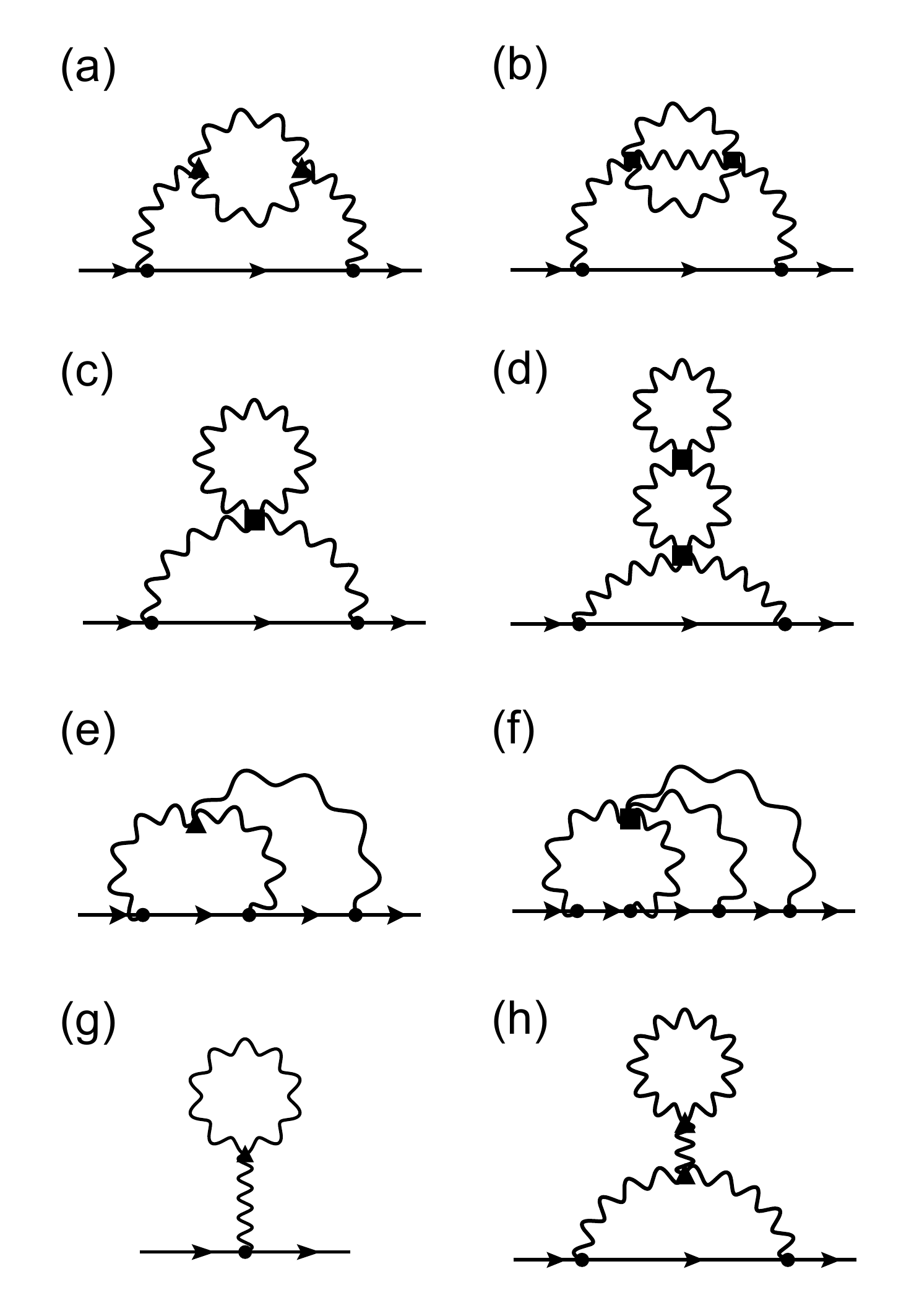}
\caption{
Low-order Feynman diagrams combining electron-phonon coupling and phonon-phonon coupling.
Each row corresponds to a correction with a specific physical meaning: 
the first row with (a) and (b) describes the three-phonon and four-phonon bubble diagrams; 
the second row with (c) and (d) describes the two lowest-order four-phonon loop diagrams modifying the phonon line in the Fan-Midgal self-energy;
the third row, panels (e) and (f), shows the vertex corrections to the electron-phonon self-energy mediated by three-phonon and four-phonon interactions, respectively;
panels (g) and (h) correspond to the three-phonon tadpole diagram contributions to the electron-phonon coupling - specifically, panel (g) modifies the electron line directly, while panel (h) modifies the phonon line in the Fan-Migdal diagram.
}
\label{fig:elahph-diagrams}
\end{figure}

%We first address the dephasing of electron-phonon coupling induced by anharmonicity, which constitutes the central focus of this work. 
The two Feynman diagrams in Figs.\,\ref{fig:elahph-diagrams}(a) and~(b) correspond to the leading order corrections to electron-phonon coupling arising from finite phonon lifetimes. Explicitly, three-phonon interactions [Fig.\,\ref{fig:elahph-diagrams}(a)] and four-phonon interactions [Fig.\,\ref{fig:elahph-diagrams}(b)] induce damping of phonons during their interaction with electrons, leading to anharmonic dephasing effects into the Fan-Migdal electron-phonon self-energy. These finite phonon lifetimes have traditionally been neglected in the study of electron-phonon interactions, and 
the key contribution of this work is the study of these effects as detailed in Sec.\,\ref{sec:anh-ph-el} below.

% Apart from this dephasing effect, we briefly highlight the other three aspects to provide a more comprehensive understanding of the interplay between phonon-phonon and electron-phonon couplings. These considerations aim to clarify the broader physical implications and underscore their relevance to the present investigation.

Anharmonic phonon-phonon interactions can also change the phonon eigenvalues and eigenstates entering electron-phonon coupling, and these effects are captured to leading order through the diagrams in Figs.\,\ref{fig:elahph-diagrams}(c) and~(d).
The most popular strategies to capture these effects do not rely on a perturbative expansion, but instead are mostly based on the non-perturbative self-consistent harmonic approximation\,\cite{Hooton1955,Errea2015} or temperature-dependent effective Hamiltonian\,\cite{Hellman2013}, in which effective phonon eigenvalues and eigenstates that incorporate the role of anharmonic phonon-phonon interactions are calculated. The standard harmonic phonons are then replaced by these effective phonons in the electron-phonon coupling calculation. These corrections are particularly important in systems where anharmonicity leads to phonon hardening that stabilizes otherwise dynamically unstable structures\,\cite{Errea2015}. 

The Feynman diagrams in Figs.\,\ref{fig:elahph-diagrams}(e) and~(f) represent the phonon-phonon-coupling-mediated vertex corrections to the electron-phonon coupling. These vertex corrections renormalize the electron-phonon coupling strength, and can in principle enhance or suppress it depending on the momentum and frequency of the exchanged phonons. Furthermore, these terms impart momentum and frequency dependence to the interaction, resulting in a non-local electron-phonon coupling. A systematic study of these effects is beyond the scope of the present work, but would be an interesting avenue of future research.
%Therefore, a systematic inclusion of such effects remains a topic for future theoretical and computational development.
%This class of diagram can become significant in systems with strong anharmonicity, and may lead to both quantitative and qualitative changes in the electron self-energy.

The tadpole diagrams presented in Figs.\,\ref{fig:elahph-diagrams}(g) and~(h) are associated with lattice distortions.
The challenge in describing these diagrams lies in the fact that most approaches rely on phonon models with fixed equilibrium positions. 
When the tadpole diagram directly acts on the electron line, as in Fig.\,\ref{fig:elahph-diagrams}(g), it corresponds to an electron subject to the thermal fluctuations driven by three-phonon coupling. Asymmetric anharmonicity can lead to thermal averages of the phonon field that do not vanish, thereby modifying the potential experienced by the electrons. If the non-vanishing vibrational average acts on the phonon line as in Fig.\,\ref{fig:elahph-diagrams}(h), this renormalises the phonons themselves.
Several works incorporate the lattice distortions associated with these diagrams, highlighting Ref.\,\cite{Monacelli2021}, which uses a quantum position operator describing the phonon model with a Gaussian distribution but without fixing the equilibrium positions to the reference lattice and instead treating them as variational parameters;
and Ref.\,\cite{Lafuente2022PRB}, which retains the coordinate operator for the Green's function describing the lattice vibrations rather than using a phonon field operator.

\subsection{Anharmonic Dephasing in the Electron-Phonon Interaction}
\label{sec:anh-ph-el}

\subsubsection{General overview}

The main objective of this work is the incorporation of anharmonic dephasing (or, equivalently, finite phonon lifetimes) into electron-phonon interactions. These effects could be incorporated using different strategies:
(i) introducing an energy linewidth to the phonon energy terms in the expression for the electron self-energy; 
(ii) using a phonon spectral function arising from the three-phonon bubble self-energy to describe phonon lines with both energy shifts and broadening; and
(iii) performing a Dyson series perturbative expansion with specific diagrams included.

The first strategy, adding an imaginary part to the phonon energy, is conceptually and computationally simple and is detailed in the Supplemental Material (SM)\,\cite{SM}. %\,\ref{si:ph_broaden}.
%can be seen as a mode-resolved broadening within the first-order electron-phonon self-energy, approximately treating dephasing as a perturbation only relevant when it exceeds intrinsic quantum fluctuations, i.e., manually given broadening parameter.
%The exact formula of this treatment is given in SI.\,\ref{si:ph_broaden}.
However, when the phonon-phonon bubble diagrams are inserted into the electron-phonon self-energy as in Figs.\,\ref{fig:elahph-diagrams}(a) and~(b), the real and imaginary parts are inherently coupled. 
As a result, simple broadening can obscure the interplay between electron-phonon coupling and phonon dephasing processes, potentially omitting critical interference or feedback effects.
%While in systems of particular interest, this intersection can play a crucial role in modifying the electron-phonon interaction, which probably makes this approximation inappropriate.

The second strategy, incorporating anharmonicity-induced dephasing through the phonon spectral function, is also detailed in the SM. %and represents a promising direction for future investigation. 
In this approach, the free-phonon propagators are replaced by interacting propagators in the spectral representation, accounting for both the real and imaginary parts of the phonon self-energy. 
%We give the related formulae in the SI.\,\ref{si:spectral_func}.
In strongly anharmonic systems the spectral broadening can be large, leading to a frequency-dependent electron-phonon coupling strength, which in turn requires frequency-resolved phonon states and potential derivatives -- quantities that are not currently accessible in practical calculations. While this is a promising research direction, we leave it for future work.

%But for now, the validity of this method becomes questionable in the strongly anharmonic regime, where significant spectral broadening necessitates full frequency integration. This implies  Furthermore, expressing the potential derivatives in the phonon basis becomes ill-defined when the quasiparticle picture breaks down.

Therefore, in this paper, we adopt the third strategy: a Dyson perturbative expansion to investigate anharmonic phonon dephasing in the context of electron-phonon coupling. One advantage of this method is that it provides a clear view of the underlying mechanisms and patterns, without relying on the quasi-particle approximation. However, a limitation is that it does not allow us to include many diagrams simultaneously due to the computational cost. Our goal here is to take an initial step towards exploring this relatively under-investigated topic.

% Now, we begin to derive the formula of describing the anharmonicity-dephasing electron-phonon scattering rates.
% The problem of the coupling of electrons with anharmonic phonons can be partially explained by the electron coupling with a phonon modified by the anharmonic multi-phonon coupling effects.
% We take the 1st-order el-ph coupling and the lowest order phonon-phonon scattering, i.e., three-phonon scattering (3-ph), and then the system's Hamiltonian has an interacting term that can be written as

\subsubsection{Derivation of anharmonic dephasing in the electron-phonon interaction}
\label{subsec:derive}

%by considering the Feynman diagram in Fig.\,\ref{fig:elahph-diagrams}, repeated in Fig.\,\ref{fig:main-diagram} with the labels associated with all degrees of freedom.
%In our case, the el-ph interaction with anharmonic phonons can be approximated by coupling electrons to phonons renormalized by multi-phonon interactions. 

In this section, we derive the formula for incorporating anharmonic dephasing in electron-phonon scattering rates. We consider a Hamiltonian incorporating first-order electron-phonon coupling and lowest-order three-phonon coupling: 
\begin{equation}
\label{eq:ep-3p}
\begin{split}
\hat{H}_{\mathrm{ep}+\mathrm{3p}} &= \hat{H}_0 +  
\frac{1}{\sqrt{N}}
\sum_{\mu_1\mu_2\lambda}
g_{\mu_2\mu_1\lambda}^{{\mathbf{k}}{\mathbf{q}}}
\hat{A}_{\lambda{\mathbf{q}}}
\hat{c}_{\mu_2{\mathbf{k}}+{\mathbf{q}}}^\dagger
\hat{c}_{\mu_1{\mathbf{k}}} \\
&+
\frac{1}{3!\sqrt{N}}
\sum_{\lambda_1\lambda_2\lambda_3}
\phi_{\lambda_1\lambda_2\lambda_3}^{{\mathbf{q}}_1{\mathbf{q}}_2{\mathbf{q}}_3}
\hat{A}_{\lambda_1{\mathbf{q}}_1}
\hat{A}_{\lambda_2{\mathbf{q}}_2}
\hat{A}_{\lambda_3{\mathbf{q}}_3},
\end{split}
\end{equation}
where $g_{\mu_2\mu_1\lambda}^{{\mathbf{k}}{\mathbf{q}}}$ are the electron-phonon coupling coefficients, $\phi_{\lambda_1\lambda_2\lambda_3}^{{\mathbf{q}}_1{\mathbf{q}}_2{\mathbf{q}}_3}$ are the 3-phonon coupling coefficients, and $\hat{A}_{\lambda{\mathbf{q}}}$ and $\hat{c}_{\mu{\mathbf{k}}}$ are the field operators for phonons and electrons, respectively.
%In this complex system, we can derive all types of Feynman diagrams, i.e., different types of physical processes, based on the perturbation technique as we mentioned earlier.

\begin{figure}
\centering
\includegraphics[width=0.65\linewidth]{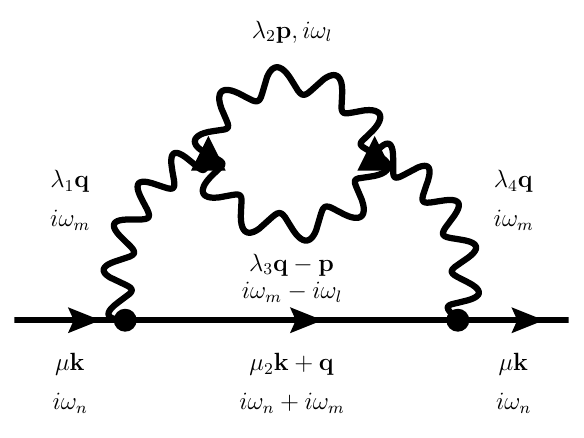}
\caption{Feynman diagram representing the three-phonon coupling mediated dephasing of the electron-phonon interaction.}
\label{fig:main-diagram}
\end{figure}

Using the same strategy as that used in Sec.\,\ref{sec:el-ph} for electron-phonon interactions and in Sec.\,\ref{sec:ph-ph} for phonon-phonon interactions, we can approximate the electron self-energy arising from the Hamiltonian in Eq.\,(\ref{eq:ep-3p}) via a perturbative expansion. As discussed in Sec.\,\ref{sec:elph-anh-overview}, we focus on the term associated with the Feynman diagram in Fig.\,\ref{fig:elahph-diagrams}(a), which we repeat in Fig.\,\ref{fig:main-diagram} including the labels associated with all degrees of freedom. From this diagram, we can directly write the self-energy as:
\begin{equation}
\begin{split}
\Sigma_{\mu{\mathbf{k}}}&(i\omega_n)= -
\frac{S}{36N^2}\Big(\frac{1}{\beta}\Big)^2
\sum_{\lambda_1\lambda_2\lambda_3\lambda_4}
\sum_{{\mathbf{q}}{\mathbf{p}}}
\sum_{\mu_1} \\
&g_{\mu_1 \mu\lambda_1}^{{\mathbf{k}}{\mathbf{q}}}
g_{\mu_1 \mu\lambda_4}^{{\mathbf{k}}{\mathbf{q}}*}
\phi_{\lambda_1\lambda_2\lambda_3}^{{\mathbf{q}}{\mathbf{p}}({\mathbf{q}}-{\mathbf{p}})*}
\phi_{\lambda_2\lambda_3\lambda_4}^{{\mathbf{p}}({\mathbf{q}}-{\mathbf{p}}){\mathbf{q}}} 
\sum_{ml}
f(i\omega_m,i\omega_l),
\end{split}
\end{equation}
where $S=18$ is the symmetry factor arising from the exchange of the three-phonon vertices, and $f(i\omega_m,i\omega_l)$ is given by:
\begin{equation}
\begin{split}
f(i\omega_m,&i\omega_l) = 
G^0_{\mu_1{\mathbf{k}}+{\mathbf{q}}}(i\omega_n-i\omega_{m})
D^0_{\lambda_1{\mathbf{q}}}(i\omega_m) \\
&D^0_{\lambda_2{\mathbf{p}}}(i\omega_l)
D^0_{\lambda_3{\mathbf{q}}-{\mathbf{p}}}(i\omega_m-i\omega_l)
D^0_{\lambda_4{\mathbf{q}}}(i\omega_m),
\end{split}
\end{equation}
where $i\omega_m$ and $i\omega_l$ are the boson poles at $i2m\pi/\beta$ and $i2l\pi/\beta$ respectively, and $i\omega_n$ corresponds to the fermion poles at $i(2n+1)\pi/\beta$.

To evaluate $\sum_{ml}f(i\omega_m,i\omega_l)$, we perform the summation in two steps. 
First, we sum over index $l$ using the relation $\sum_{l}f(i\omega_m,i\omega_l)=-\beta\sum_{z}^{f(i\omega_m,z)} \mathrm{Re}\{{f(i\omega_m,z)}n_\mathrm{B}(z)\}$, while treating the $i\omega_m$ as a constant with respect to the $l$ summation.
The detailed expression is given below, and we merge the momentum index into the mode index for the derivation in this section:
\begin{equation}
\begin{split}
f(z) = A_{i\omega_m}
&\left(\frac{1}{z-\omega_{\lambda_2}} - 
\frac{1}{z+\omega_{\lambda_2}}\right) \\
&\left(\frac{1}{i\omega_m-z-\omega_{\lambda_3}} - 
\frac{1}{i\omega_m-z+\omega_{\lambda_3}}\right),
\end{split}
\end{equation}
which possesses four poles $z_1=\omega_{\lambda_2}$, $z_2=-\omega_{\lambda_2}$, $z_3=i\omega_m-\omega_{\lambda_3}$, and $z_4=i\omega_m+\omega_{\lambda_3}$ and the pure $m$-index term $A_{i\omega_m}$ is given by:
\begin{equation}
A_{i\omega_m} = 
    G^0_{\mu_1{\mathbf{k}}+{\mathbf{q}}}(i\omega_n-i\omega_m)
    D^0_{\lambda_1{\mathbf{q}}}(i\omega_m)
    D^0_{\lambda_4{\mathbf{q}}}(i\omega_m).
\end{equation}
We can thus simplify the original expression into:
\begin{equation}
\begin{split}
\sum_{i\omega_l}\frac{f(i\omega_m,i\omega_l)}{-\beta}=
A_{i\omega_m}& \\
\left\{-
\frac{\mathcal{N}_{\lambda_2}+\mathcal{N}_{\lambda_3}+1}
{i\omega_{m}-\omega_{\lambda_2}-\omega_{\lambda_3}}\right.
&+
\frac{\mathcal{N}_{\lambda_2}-\mathcal{N}_{\lambda_3}}
{i\omega_{m}-\omega_{\lambda_2}+\omega_{\lambda_3}} \\
-
\frac{\mathcal{N}_{\lambda_2}-\mathcal{N}_{\lambda_3}}
{i\omega_{m}+\omega_{\lambda_2}-\omega_{\lambda_3}}
&+
\left.\frac{\mathcal{N}_{\lambda_2}+\mathcal{N}_{\lambda_3}+1}
{i\omega_{m}+\omega_{\lambda_2}+\omega_{\lambda_3}}
\right\},
\end{split}
\end{equation}
where the denominator of each term corresponds to a distinct pattern for the 3-phonon scattering process, with minus and plus signs representing emission and absorption processes, respectively. 
We label these terms according to the signs preceding $\omega_{\lambda_2}$ and $\omega_{\lambda_3}$, for example $h^{--}$ corresponds to the term with the denominator $i\omega_{m} - \omega_{\lambda_2} - \omega_{\lambda_3}$; and these terms are explicitly given by:
\begin{widetext}
\begin{equation}
%\begin{split}
h^{--}(i\omega_m) = 
\frac{1}{i\omega_n-i\omega_m-\varepsilon_{\mu_1}}
\left(\frac{1}{i\omega_m-\omega_{\lambda_1}}-
\frac{1}{i\omega_m+\omega_{\lambda_1}}\right) %\\
%&\times
\left(\frac{1}{i\omega_m-\omega_{\lambda_4}}-
\frac{1}{i\omega_m+\omega_{\lambda_4}}\right)
\frac{\mathcal{N}_{\lambda_2}+\mathcal{N}_{\lambda_3}+1}
{i\omega_{m}-\omega_{\lambda_2}-\omega_{\lambda_3}},
%\end{split}
\end{equation}

\begin{equation}
%\begin{split}
h^{-+}(i\omega_m) = 
\frac{1}{i\omega_n-i\omega_m-\varepsilon_{\mu_1}}
\left(\frac{1}{i\omega_m-\omega_{\lambda_1}}-
\frac{1}{i\omega_m+\omega_{\lambda_1}}\right) %\\
%&\times
\left(\frac{1}{i\omega_m-\omega_{\lambda_4}}-
\frac{1}{i\omega_m+\omega_{\lambda_4}}\right)
\frac{\mathcal{N}_{\lambda_2}-\mathcal{N}_{\lambda_3}}
{i\omega_{m}-\omega_{\lambda_2}+\omega_{\lambda_3}},
%\end{split}
\end{equation}

\begin{equation}
%\begin{split}
h^{+-}(i\omega_m) = 
\frac{1}{i\omega_n-i\omega_m-\varepsilon_{\mu_1}}
\left(\frac{1}{i\omega_m-\omega_{\lambda_1}}-
\frac{1}{i\omega_m+\omega_{\lambda_1}}\right) %\\
%&\times
\left(\frac{1}{i\omega_m-\omega_{\lambda_4}}-
\frac{1}{i\omega_m+\omega_{\lambda_4}}\right)
\frac{\mathcal{N}_{\lambda_2}-\mathcal{N}_{\lambda_3}}
{i\omega_{m}+\omega_{\lambda_2}-\omega_{\lambda_3}},
%\end{split}
\end{equation}

\begin{equation}
%\begin{split}
h^{++}(i\omega_m) = 
\frac{1}{i\omega_n-i\omega_m-\varepsilon_{\mu_1}}
\left(\frac{1}{i\omega_m-\omega_{\lambda_1}}-
\frac{1}{i\omega_m+\omega_{\lambda_1}}\right) %\\
%&\times
\left(\frac{1}{i\omega_m-\omega_{\lambda_4}}-
\frac{1}{i\omega_m+\omega_{\lambda_4}}\right)
\frac{\mathcal{N}_{\lambda_2}+\mathcal{N}_{\lambda_3}+1}
{i\omega_{m}+\omega_{\lambda_2}+\omega_{\lambda_3}}.
%\end{split}
\end{equation}
\end{widetext}
Based on this notation, the remaining summation over $m$ can be written succinctly as:
\begin{equation}
\begin{split}
\sum_{m}\frac{f(i\omega_m)}{(\beta)^2}=&
-
\sum_{z\ of\ h^{--}}\mathrm{Re}\left\{h^{--}(z)\right\}\cdot n_\mathrm{B}(z) \\
&+\sum_{z\ of\ h^{-+}}\mathrm{Re}\left\{h^{-+}(z)\right\}\cdot n_\mathrm{B}(z) \\
&
-
\sum_{z\ of\ h^{+-}}\mathrm{Re}\left\{h^{+-}(z)\right\}\cdot n_\mathrm{B}(z) \\
&+\sum_{z\ of\ h^{++}}\mathrm{Re}\left\{h^{++}(z)\right\}\cdot n_\mathrm{B}(z).
\end{split}
\end{equation}
The remaining summation over $m$ is lengthy but straightforward, and it is detailed in the SM.
%At this stage, we note that the pole structure can be different depending on whether $\lambda_1$ is equal or different to $\lambda_4$.
%Therefore, we split our discussion into two parts depending on whether $\lambda_1=\lambda_4$ or $\lambda_1\neq\lambda_4$, and the detailed summation process is described in the SM.

Upon evaluation of the $m$ summation, we obtain the key result of this work, the expression for the scattering rate associated with anharmonic dephasing on the electron-phonon interaction: 
\begin{equation}
\Gamma_{\mu\mathbf{k}}^{\text{el-ah-ph}} = 
2\Gamma_{\mu\mathbf{k}}^{(\mathrm{1e1a})} + 
\Gamma_{\mu\mathbf{k}}^{(\mathrm{2e})} + 
\Gamma_{\mu\mathbf{k}}^{(\mathrm{2a})}.
\label{eq:main-equation}
\end{equation}
The one phonon emitted and one phonon absorbed term (1e1a) can be written as:
\begin{widetext}
\begin{equation}
\begin{split}
\Gamma_{\mu\mathbf{k}}^{(\mathrm{1e1a})} = 
\frac{\pi}{N^2} 
\sum_{\lambda_2\lambda_3}
\sum_{\mathbf{q}\mathbf{p}}
&\left| 
    \sum_{\lambda_1} 
    \left(
        \frac{
            g_{\mu_1\mu\lambda_1}^{{\mathbf{k}}{\mathbf{q}}}
            \phi_{\lambda_1\lambda_2\lambda_3}^{{\mathbf{q}}{\mathbf{p}}({\mathbf{q}}-{\mathbf{p}})*}
        }{
            \omega_{\lambda_2}-\omega_{\lambda_3}-\omega_{\lambda_1}+i\eta
        }
        -
        \frac{
            g_{\mu_1\mu\lambda_1}^{{\mathbf{k}}{\mathbf{q}}}
            \phi_{\lambda_1\lambda_2\lambda_3}^{{\mathbf{q}}{\mathbf{p}}({\mathbf{q}}-{\mathbf{p}})*}
        }{
            \omega_{\lambda_2}-\omega_{\lambda_3}+\omega_{\lambda_1}+i\eta
        }
    \right)
\right|^2  \\
&\times
\delta({\omega_n} - \omega_{\lambda_2} + \omega_{\lambda_3} - \varepsilon_{\mu_1})
\left[
\mathcal{N}_{\lambda_2}\mathcal{N}_{\lambda_3} + \mathcal{N}_{\lambda_2}
\right],
\end{split}
\label{eq:gamma-1e1a}
\end{equation}
%\end{widetext}
the two phonon emitted term (2e) can be written as:
%\begin{widetext}
\begin{equation}
\begin{split}
\Gamma_{\mu\mathbf{k}}^{(\mathrm{2e})} = 
\frac{\pi}{N^2} 
\sum_{\lambda_2\lambda_3}
\sum_{\mathbf{q}\mathbf{p}}
&\left| 
    \sum_{\lambda_1} 
    \left(
        \frac{
            g_{\mu_1\mu\lambda_1}^{{\mathbf{k}}{\mathbf{q}}}
            \phi_{\lambda_1\lambda_2\lambda_3}^{{\mathbf{q}}{\mathbf{p}}({\mathbf{q}}-{\mathbf{p}})*}
        }{
            \omega_{\lambda_2}+\omega_{\lambda_3}-\omega_{\lambda_1}+i\eta
        }
        -
        \frac{
            g_{\mu_1\mu\lambda_1}^{{\mathbf{k}}{\mathbf{q}}}
            \phi_{\lambda_1\lambda_2\lambda_3}^{{\mathbf{q}}{\mathbf{p}}({\mathbf{q}}-{\mathbf{p}})*}
        }{
            \omega_{\lambda_2}+\omega_{\lambda_3}+\omega_{\lambda_1}+i\eta
        }
    \right)
\right|^2  \\
&\times\delta({\omega_n} - \omega_{\lambda_2} - \omega_{\lambda_3} - \varepsilon_{\mu_1})
\left[
(\mathcal{N}_{\lambda_2}+\mathcal{N}_{\lambda_3}+1)(f_{\mu_1}-1) + \mathcal{N}_{\lambda_2}\mathcal{N}_{\lambda_3}
\right],
\end{split}
\label{eq:gamma-2e}
\end{equation}
%\end{widetext}
and the two phonon absorbed term (2a) can be written as:
%\begin{widetext}
\begin{equation}
\begin{split}
\Gamma_{\mu \mathbf{k}}^{(\mathrm{2a})} = 
-\frac{\pi}{N^2} 
\sum_{\lambda_2\lambda_3}
\sum_{\mathbf{q}\mathbf{p}}
&\left| 
\sum_{\lambda_1} 
\left(
    \frac{
        g_{\mu_1\mu\lambda_1}^{{\mathbf{k}}{\mathbf{q}}}
        \phi_{\lambda_1\lambda_2\lambda_3}^{{\mathbf{q}}{\mathbf{p}}({\mathbf{q}}-{\mathbf{p}})*}
    }{
        \omega_{\lambda_2}+\omega_{\lambda_3}-\omega_{\lambda_1}+i\eta
    }
    -
    \frac{
        g_{\mu_1\mu\lambda_1}^{\mathbf{k}{\mathbf{q}}}
        \phi_{\lambda_1\lambda_2\lambda_3}^{{\mathbf{\mathbf{q}}}{\mathbf{p}}({\mathbf{\mathbf{q}}}-{\mathbf{p}})*}
    }{
        \omega_{\lambda_2}+\omega_{\lambda_3}+\omega_{\lambda_1}+i\eta
    }
\right)
\right|^2  \\
\times\delta({\omega_n} + \omega_{\lambda_2} + &\omega_{\lambda_3} - \varepsilon_{\mu_1})
\left[
(\mathcal{N}_{\lambda_2}+\mathcal{N}_{\lambda_3}+1)(f_{\mu_1}-1) +
\mathcal{N}_{\lambda_2}\mathcal{N}_{\lambda_3} + \mathcal{N}_{\lambda_2} + \mathcal{N}_{\lambda_3} + 1
\right].
\label{eq:gamma-2a}
\end{split}
\end{equation}
\end{widetext}

\section{First principles implementation}
\label{sec:comput}

\begin{figure*}[t]
    \centering
    \includegraphics[width=\linewidth]{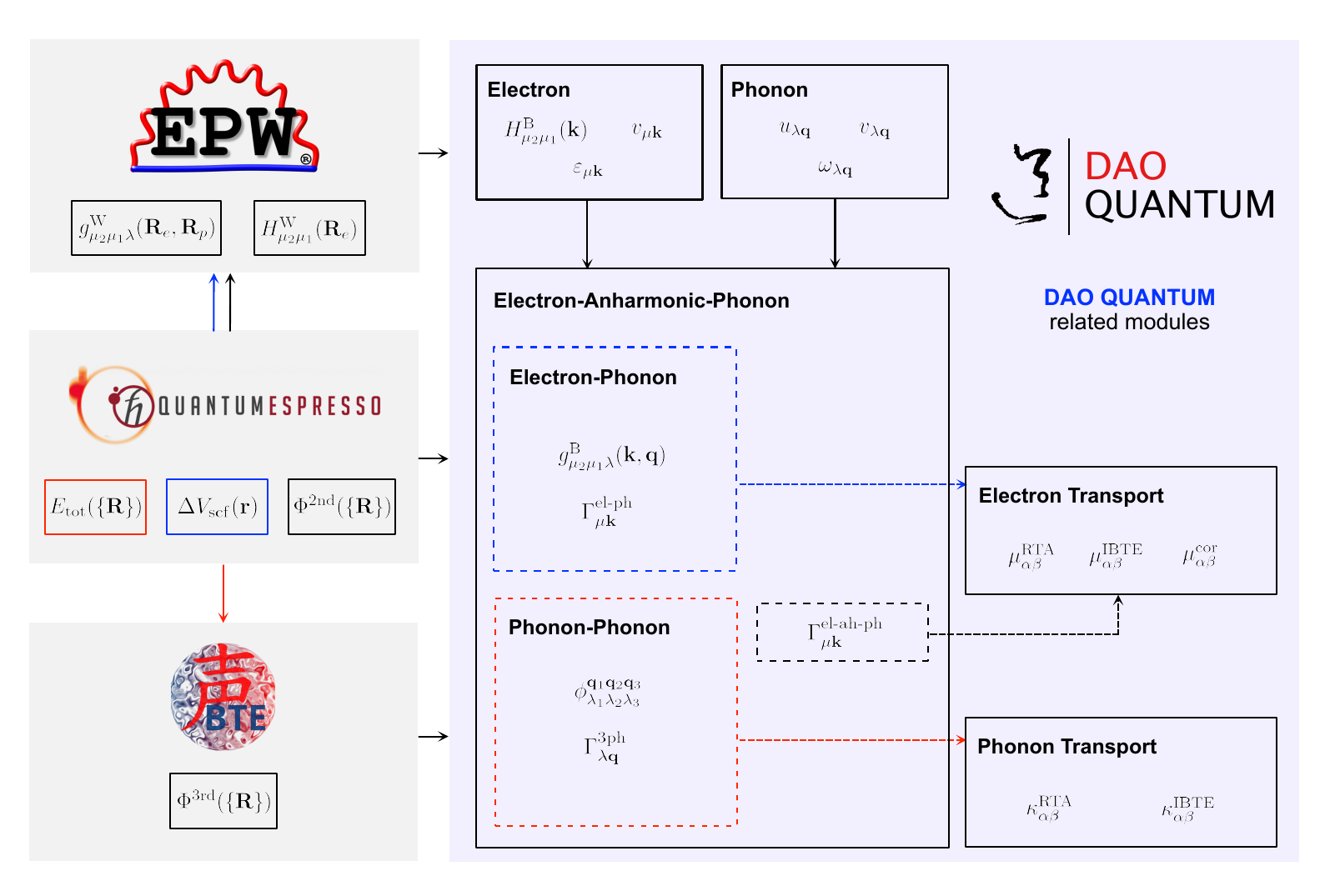}
    \caption{Workflow for the first principles implementation of the electron-phonon and electron-anharmonic-phonon calculations. The blue box contains the quantities associated with electron-phonon coupling, the red box contains the quantities associated with phonon-phonon coupling, and the black color contains the quantities associated with electron-anharmonic-phonon coupling.}
    \label{fig:workflow}
\end{figure*}

The theoretical description of anharmonic dephasing in electron-phonon coupling is provided above in Sec.\,\ref{sec:theory}, culminating in the Feynman diagram in Fig.\,\ref{fig:main-diagram} and the associated Eqs.\,(\ref{eq:main-equation})-(\ref{eq:gamma-2a}) for the scattering rate. In this section, we present a first principles implementation of anharmonic dephasing in electron-phonon coupling, which is schematically summarized in Fig.\,\ref{fig:workflow}. 
%The calculation is feasible but computationally expensive, so we also put forward a computationally simpler heuristic index to measure the strength of anharmonic corrections, allowing for a quick assessment of whether the inclusion of anharmonic effects using a full calculation is warranted.

\subsection{Electron, phonon, and electron-phonon parameters}

The key quantities required for the first principles evaluation of anharmonic dephasing in electron-phonon coupling involve electrons, phonons, and their coupling.

For electrons, the required information is captured by the electronic Hamiltonian in the Bloch gauge $H^{\mathrm{B}}_{\mu_2\mu_1}(\mathbf{k})$, and the associated single-particle band energies $\mathcal{\varepsilon}_{\mu\mathbf{k}}$ and group velocities $v_{\mu\mathbf{k}}$. 

For phonons, the required information is captured by the second-order interatomic force constants $\Phi^{\mathrm{2nd}}(\{\mathbf{R}\})$ and corresponding dynamical matrix $\phi_{\kappa_1\kappa_2}^{\alpha_1\alpha_2}(\mathbf{q})$, and the associated phonon frequencies $\mathcal{\omega}_{\lambda\mathbf{q}}$, eigenvectors $u_{\lambda\mathbf{q}}$, and group velocities $v_{\lambda\mathbf{q}}$. The anharmonic phonon-phonon interactions are encoded by the third-order interatomic force constants $\Phi^{\mathrm{3rd}}(\{\mathbf{R}\})$ and the associated three-phonon matrix elements $\phi_{\lambda_1\lambda_2\lambda_3}^{\mathbf{q}_1\mathbf{q}_2\mathbf{q}_3}$ and linewidths $\Gamma^{3\mathrm{ph}}_{\lambda\mathbf{q}}$.

The electron-phonon interaction is characterized by the coupling matrix elements in the Bloch gauge $g^\mathrm{B}_{\mu_2\mu_1\lambda}(\mathbf{k}, \mathbf{q})$ and the corresponding scattering rates $\Gamma^{\text{el-ph}}_{\mu\mathbf{k}}$. 

Bringing all of these together, we can construct the anharmonic electron-phonon processes described in Eqs.\,(\ref{eq:gamma-1e1a})--(\ref{eq:gamma-2a}), providing a concrete strategy for calculating the scattering rate $\Gamma^{\text{el-ah-ph}}_{\mu\mathbf{k}}$ in Eq.\,(\ref{eq:main-equation}).
% The key quantities are the matrix elements $\mathcal{M}^{\text{el-ah-ph}}_{\mu_1\mu_2\lambda}(\mathbf{k}, \mathbf{p}, \mathbf{q})$, contributing to the total scattering rate $\Gamma^{\mathrm{el-ah-ph}}_{\mu_1\mathbf{k}}$.

More generally, the workflow in Fig.\,\ref{fig:workflow} also shows how the evaluation of anharmonic electron-phonon scattering rates can then be used as a building block to evaluate electron transport incorporating anharmonic dephasing into the mobility $\mu_{\alpha\beta}$. While the focus of this work are the anharmonic electron-phonon scattering rates, we report a calculation of electron mobility in MgB$_2$ incorporating anharmonic dephasing in an accompanying work. 
To highlight this connection, we present the standard formulas that relate the scattering rates to the electron mobility and electrical conductivity within the relaxation-time approximation as follow
\begin{equation}
\sigma_{\alpha\beta}=
\frac{e}{n_c V_{\mathrm{uc}}}
\sum_{n\mathbf{k}}
v_{\mu\mathbf{k}}^\alpha\,v_{\mu\mathbf{k}}^\beta\,
\tau_{\mu\mathbf{k}}
\left(-\frac{\partial f^{0}_{\mu\mathbf{k}}}{\partial \varepsilon_{\mu\mathbf{k}}}\right),
\end{equation}
with the electrical conductivity given by
\begin{equation}
\mu_{\alpha\beta}=\frac{\sigma_{\alpha\beta}}{n_c e}.
\end{equation}
For completeness, we also highlight that the workflow in Fig.\,\ref{fig:workflow} can naturally be used to capture phonon transport in terms of the thermal conductivity $\kappa_{\alpha\beta}$, but this will be reported elsewhere.

\subsection{Workflow}

The first principles implementation relies on our in-house code \texttt{DaoQuantum}, which will be fully described elsewhere. 
In short, \texttt{DaoQuantum} is developed in C++ (with a Python interface) with a design that enables heterogeneous computation across different hardware architectures, including GPUs. It is currently integrated into a workflow with several existing first principles packages as schematically shown in Fig.\,\ref{fig:workflow}.

We use \texttt{Quantum ESPRESSO}\,\cite{Giannozzi2009} for both DFT and DFPT calculations to describe non-interacting electrons and phonons. The Kohn-Sham potential derivatives $\Delta V_\mathrm{scf}(\mathbf{r})$ contributing to the electron-phonon coupling coefficients are directly obtained from DFPT, while the long-range ionic contribution is computed using \texttt{EPW}\,\cite{Lee2023}. The sum of these two components is then rotated to $\mathbf{q}$ points sampled across the full Brillouin zone.
Maximally localized Wannier functions (MLWF)\,\cite{Marzari2012} generated by \texttt{Wannier90}\,\cite{Mostofi2014} are used to interpolate the matrix elements on a finer mesh in reciprocal space\,\cite{Giustino2007}.
We adopt the finite displacement method to obtain the third-order force constants using the \texttt{thirdorder.py} script within the \texttt{ShengBTE} package\,\cite{Li2014}. It extracts the total energies $E_\mathrm{tot}(\{\mathbf{R}\})$ and atomic forces calculated by \texttt{Quantum ESPRESSO} and computes the force constants by numerically differentiating the forces from displaced supercell configurations.

These physical quantities are then fed to \texttt{DaoQuantum} for the evaluation of anharmonic dephasing in electron-phonon coupling.
Our workflow starts from the electron Hamiltonian $H^\mathrm{W}_{\mu_2\mu_1}(\mathbf{R}_\mathrm{e})$ and electron-phonon coupling matrix $g^\mathrm{W}_{\mu_2\mu_1\lambda}(\mathbf{R}_\mathrm{e},\mathbf{R}_\mathrm{p})$ in the Wannier representation, imported from \texttt{EPW}\,\cite{Lee2023}. A Fourier transform enables efficient interpolation onto dense $\mathbf{k}$-grids, providing eigenvalues and group velocities while preserving the gauge structure encoded in the eigenvectors.
This gauge information is crucial for transforming physical quantities between the MLWF gauge and the Bloch gauge.
The MLWF gauge exploits the fact that smoothness in reciprocal space leads to spatial localization in real space, requiring $\nabla_{\mathbf{k}} \ket{u_{\mu\mathbf{k}}}$ to be well-defined. This localization, in turn, enables an accurate Fourier interpolation, which underlies the MLWF interpolation method.
For phonons, we use the second-order force constants as input. From these, the phonon eigenvalues, eigenvectors, and group velocities are computed throughout the Brillouin zone.

In our implementation, the electron-phonon coupling module transforms the coupling coefficients from the Wannier to the Bloch representation on a dense $\mathbf{k}$-grid. The resulting interpolated matrix elements are then employed to compute the Fan–Migdal self-energy.
Phonon-phonon coupling, specifically the three-phonon interaction, is evaluated by contracting the Fourier-transformed third-order force constants with phonon eigenvectors to obtain the interaction vertices.
These vertices are subsequently used to compute the bubble self-energy.
The contributions from both electron-phonon and phonon-phonon couplings are then combined to obtain the correction to the imaginary part of the electron self-energy.

A practical demonstration of this implementation is presented in Sec.\,\ref{sec:materials} below, where the anharmonic dephasing of the electron-phonon interaction is calculated for Si, SiC, and PbTe. Additionally, a companion work presents calculations for MgB$_2$ and also uses the anharmonic electron-phonon scattering rates to evaluate the impact of anharmonic dephasing on electron mobility.

\section{Anharmonic dephasing in electron-phonon coupling in materials}
\label{sec:materials}

In this section we use the theory presented in Sec.\,\ref{sec:theory} and the first principles implementation presented in Sec.\,\ref{sec:comput} to evaluate anharmonic dephasing in the electron-phonon interaction of a range of materials. 
%This part focuses on real-system applications and may offer insights into the underlying scattering mechanisms.
We select Si, SiC, and PbTe as our example materials, owing to their wide-ranging applications in electronics, optoelectronics, and thermoelectrics. These results are complemented by the calculation of conductivity in MgB$_2$ reported in an accompanying work, in which anharmonic dephasing of electron-phonon coupling is found to make a substantial contribution. 

In this section, we will compare the electron-phonon coupling properties calculated using the standard approach, harmonic phonons and linear electron-phonon coupling, with electron-phonon coupling incorporating anharmonic phonon dephasing. For simplicity, we will refer to the former as ``harmonic electron-phonon coupling'' and to the latter as ``anharmonic electron-phonon coupling''.

% \begin{figure*}[t]
%     \centering
%     \includegraphics[width=\linewidth]{application.pdf}
%     \caption{SiC (a) band structure and (b) phonon dispersion}
%     \label{fig:application}
% \end{figure*}

\subsection{Silicon}
\label{sec:si}

\begin{figure*}[t]
\centering
\includegraphics[width=\linewidth]{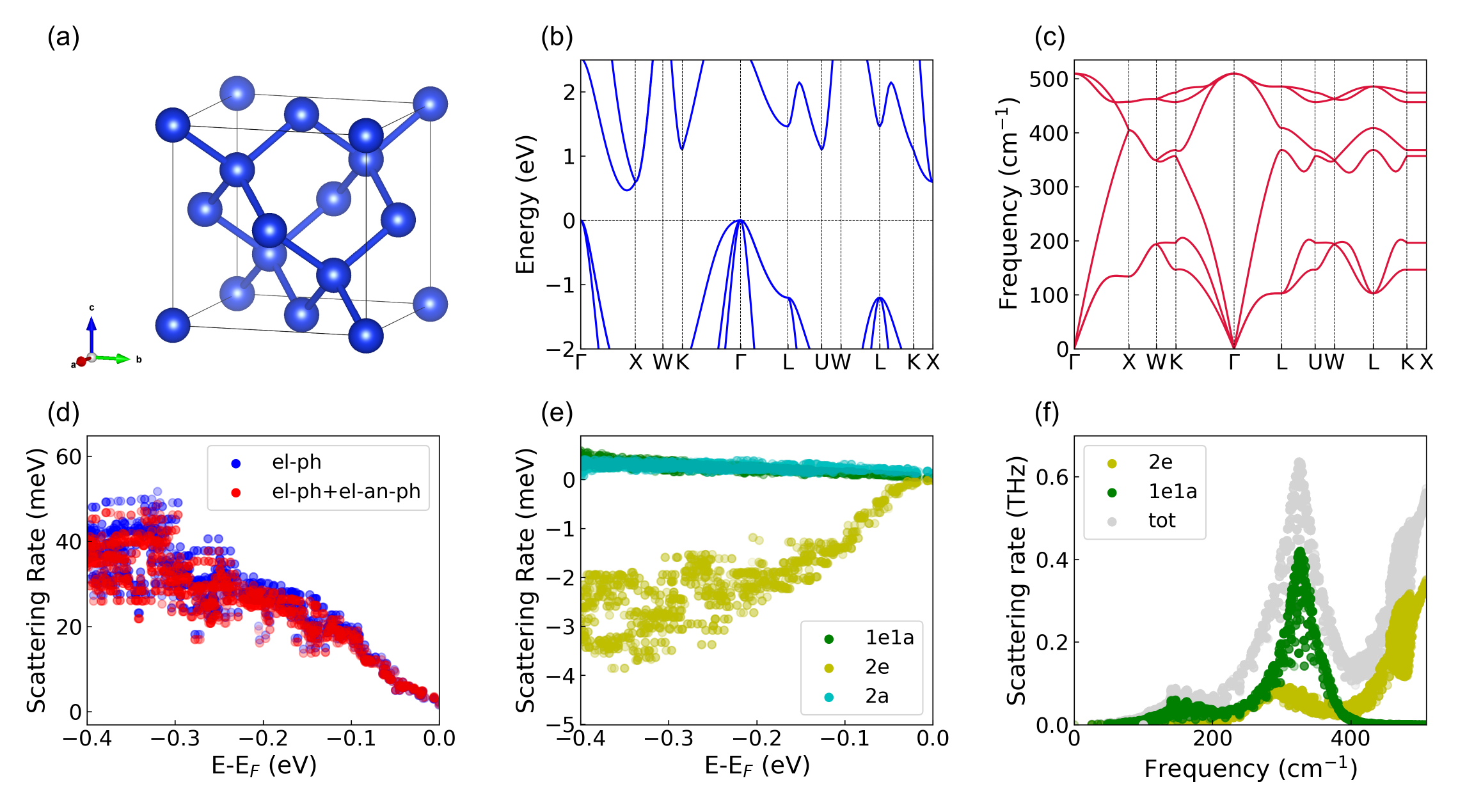}
\caption{Silicon results, including (a) crystal structure; (b) electron band structure; (c) phonon dispersion; (d) electron-phonon scattering rates in blue and electron-phonon plus electron-anharmonic-phonon scattering rates in red at $300$\,K; (e) process-resolved electron-anharmonic-phonon correction versus electron energy; (f) process-resolved three-phonon scattering rates versus phonon frequencies.}
\label{fig:si}
\end{figure*}

Silicon is a prototypical semiconductor with well-characterized electronic and phononic properties, making it a natural starting point for investigating anharmonic dephasing in electron-phonon coupling.
We perform first-principles calculations on Si using the same parameters as in Ref.\,\cite{PONCE2016}.
Its crystal structure, electron band structure, and phonon dispersion are given in Figs.\,\ref{fig:si}(a), (b), and~(c), respectively.
Its band structure shows that silicon is an indirect band gap semiconductor, with the valence band maximum at the $\Gamma$ point and the conduction band minimum located along the $\Gamma$–$X$ high symmetry line. The indirect nature of the silicon band gap leads to intravalley scattering dominating electron-phonon interactions.

Figure~\ref{fig:si}(d) shows the harmonic electron-phonon coupling scattering rates (blue circles) and the anharmonic electron-phonon coupling scattering rates (red circles) as a function of the electron energy within the valence bands. Both scattering rates increase as the energy moves deeper below the valence band maximum, as expected due to the increase in the electron density of states and the associated increase in the phase space available for electron-phonon scattering processes. The two scattering rates are relatively similar, but there is a clear decrease in the scattering rate driven by anharmonic phonon dephasing.

We show the different electron-anharmonic-phonon scattering mechanisms in Fig.\,\ref{fig:si}(e) as a function of the electron energy within the valence bands. The results indicate that the dominant scattering process is two-phonon emission (olive green circles), which provides a negative contribution to the scattering rate, explaining the overall smaller scattering rate compared to the harmonic electron-phonon coupling. 

Figure~\ref{fig:si}(f) shows the energy-conserving phonon-phonon scattering mechanisms as a function of the harmonic phonon frequency. Two-phonon emission (olive green circles) is strongest at high frequencies around $450$-$500$\,cm$^{-1}$, where a relatively flat phonon dispersion facilitates the energy and momentum conservation required for two-phonon decay. The one-phonon emission and one-phonon absorption processes (dark green circles) dominate in the 300–400\,cm$^{-1}$ range, but they make a small contribution to electron-phonon scattering near the band edge [c.f. Fig.\,\ref{fig:si}(e)], indicating that anharmonic effects are mode-dependent, shaped by phonon dephasing and their coupling to electronic states.

\subsection{Silicon Carbide}
\label{sec:sic}

\begin{figure*}[t]
\centering
\includegraphics[width=\linewidth]{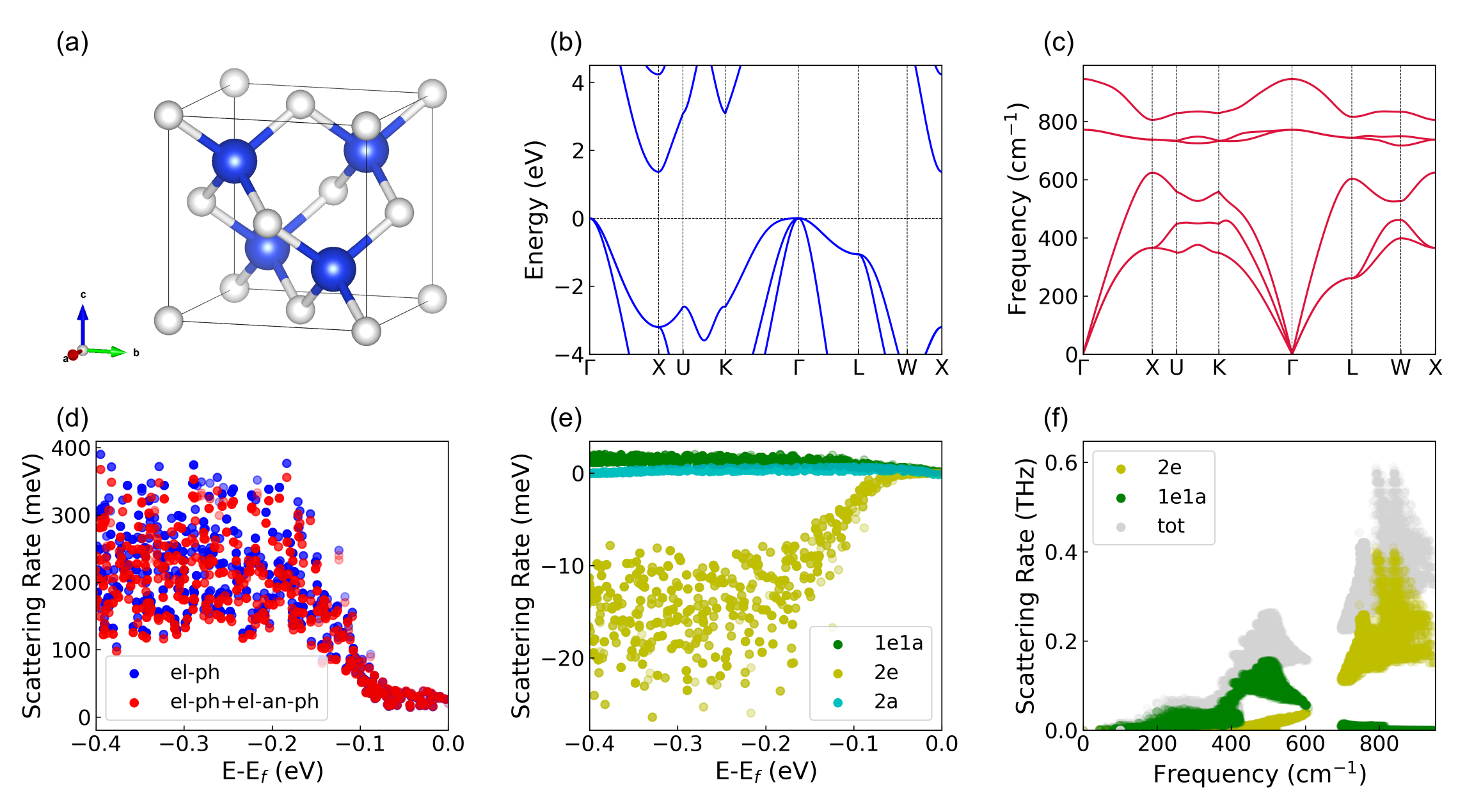}
\caption{Silicon carbide restuls, including (a) crystal structure; (b) electron band structure; (c) phonon dispersion; (d) electron-phonon scattering rates in blue and electron-phonon plus electron-anharmonic-phonon scattering rates in red at $300$\,K; (e) process-resolved electron-anharmonic-phonon correction versus electron energies; (f) process-resolved three-phonon scattering rates versus phonon frequencies.}
\label{fig:sic}
\end{figure*}

Silicon carbide, noted for its mechanical robustness and high-temperature electronic performance, has been widely studied experimentally and theoretically.
We study SiC from first principles using the same parameters as in Ref.\,\cite{NOFFSINGER2010}.
Its crystal structure, electron band structure, and phonon dispersion are given in Figs.\,\ref{fig:sic}(a), (b), and~(c), respectively.
The electronic band structure in Fig.\,\ref{fig:sic}(b) shows a relatively flat and degenerate valence band near the $\Gamma$ point.
The valley extends anisotropically from the $\Gamma$ point with varying effective masses, 
%a key factor in our analysis of hole-phonon scattering.
a feature that underpins the transport behavior of SiC, where weak intravalley processes dominate near the band edge, while intervalley scattering becomes significant away from the Fermi level beyond $0.1$\,eV, contributing predominantly to the total scattering rate\,\cite{Meng2019}.

The electron-phonon scattering rates within the harmonic (blue circles) and anharmonic (red circles) approximations are compared in Fig.\,\ref{fig:sic}(d) in the energy window around the top of the valence band. They have a similar magnitude, which grows as the energy moves deeper within the valence band due to the increase in the phase space available for scattering. The anharmonic correction is again small, and Fig.\,\ref{fig:sic}(e) shows the different electron-anharmonic-phonon scattering mechanisms as a function of the electron energy within the valence bands. The anharmonic correction grows as we move away from the band edge, indicating that more electrons are interacting with phonons whose coherence is disrupted by anharmonicity.
As discussed earlier, scattering near the band edge is dominated by weak intravalley interactions but, as we move away from the band edge, intervalley scattering becomes more significant and contributes substantially to the overall scattering rates. 
Similar to Si, the two-phonon emission process dominates anharmonic contributions in SiC, with its impact increasing rapidly alongside the growth of intervalley scattering. 
Figure~\ref{fig:sic}(f) illustrates the process-resolved phonon-phonon scattering rates plotted against the harmonic phonon frequency. 
Similar to Si, due to the relatively flat phonon dispersion at high frequencies around $800$–$850$\,cm$^{-1}$, two-phonon emission becomes particularly prominent in SiC, as it allows for efficient energy and momentum conservation in the decay process.
% Two-phonon emission (olive green circles) is strongest at high frequencies around $800$-$850$\,cm$^{-1}$, where a relatively flat phonon dispersion facilitates the energy and momentum conservation required for two-phonon decay. 
The one-phonon emission and one-phonon absorption processes dominate in the $500$–$550$\,cm$^{-1}$ range, but again
only provide a small contribution to electron-phonon scattering near the band edge [c.f. Fig.\,\ref{fig:sic}(e)].
%, further proving the point that anharmonic effects are mode-dependent, shaped by phonon dephasing and their coupling to electronic states.

\subsection{Lead telluride}
\label{sec:pbte}

\begin{figure*}[t]
\centering
\includegraphics[width=\linewidth]{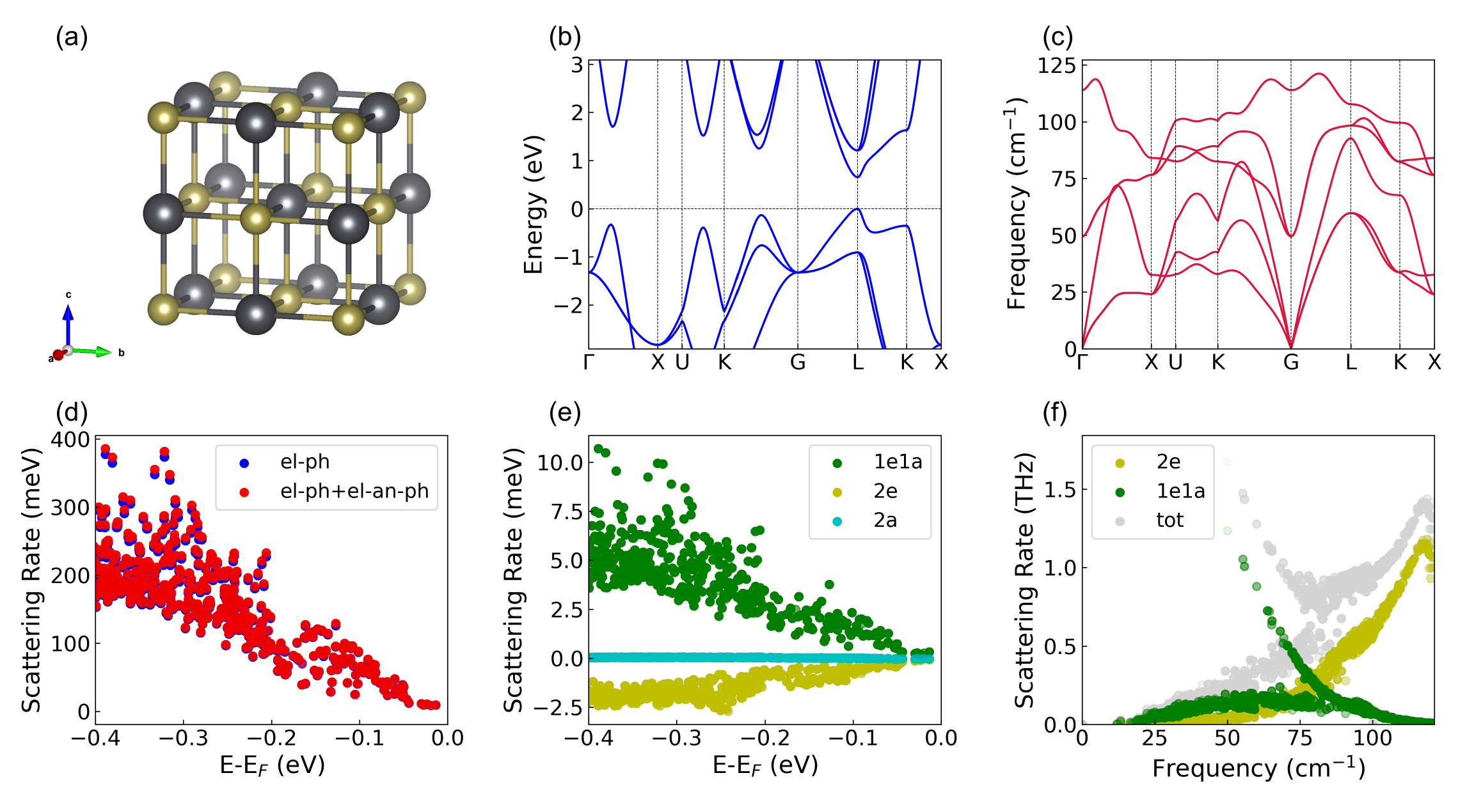}
\caption{Lead telluride results, including (a) crystal structure; (b) electron band structure; (c) phonon dispersion; (d) electron-phonon scattering rates in blue and electron-phonon plus electron-anharmonic-phonon scattering rates in red at $300$\,K; (e) process-resolved electron-anharmonic-phonon correction versus electron energies; (f) process-resolved three-phonon scattering rates versus phonon frequencies.}
\label{fig:pbte}
\end{figure*}

Lead telluride is a well-known thermoelectric material, whose crystal structure is presented in Fig.\,\ref{fig:pbte}(a). 
%Extensive studies have explored band engineering strategies to enhance its thermoelectric performance\,\cite{Zhang2012,He2019}. At the same time, 
It exhibits strong anharmonic lattice dynamics\,\cite{Delaire2011}, making it an ideal platform for investigating the interplay between electron-phonon and phonon-phonon couplings.
This is particularly relevant as high-performance thermoelectric materials typically exhibit high electrical conductivity and low thermal conductivity, and anharmonic phonon dephasing may play a significant role in both.

We perform first principles calculations for PbTe using the same parameters as in Ref.\,\cite{Cao2021}.
The electronic band structure is shown in Fig.\,\ref{fig:pbte}(b) and the phonon dispersion in Fig.\,\ref{fig:pbte}(c). PbTe exhibits a multi-valley structure where the conduction band minima and valence band maxima are located at the four equivalent $L$ points in the Brillouin zone. Moreover, multiple additional band extrema exist close to the band edge.
This multi-valley band structure is key to the favorable thermoelectric performance of PbTe as it increases the carrier density of states. It also makes intravalley scattering the dominant electron-phonon mechanism.

The electron-phonon scattering rates within the harmonic (blue circles) and anharmonic (red circles) approximations are compared in Fig.\,\ref{fig:pbte}(d) in the energy window around the top of the valence band. 
Similar to Si and SiC, the magnitudes of the scattering rates in PbTe grow as the energy moves deeper within the valence band driven by the increase in the phase space available for scattering. The anharmonic correction is again small but, unlike Si an SiC, it \textit{increases} the scattering rate in PbTe. Figure~\ref{fig:pbte}(e) shows the different electron-anharmonic-phonon scattering mechanisms as a function of the electron energy within the valence bands, indicating a dominant positive one-electron emission and one-electron absorption (1e1a) process (dark green circles), which explains the overall increase in the scattering rate upon inclusion of anharmonic dephasing. Figure~\ref{fig:pbte}(f) shows that the 1e1a scattering processes (dark green circles) occur mainly at phonon frequencies between 50\,cm$^{-1}$ and 75\,cm$^{-1}$. This mechanism has previously been reported as the resonant scattering process LA $+$ LO $\to$ TO within the valley around the $\Gamma$ point, which leads to the ``waterfall'' pattern in the three-phonon scattering\,\cite{Delaire2011}.

%\subsection{Temperature dependence}

%All exhibit a strong temperature dependence, with values increasing as temperature rises.
%PbTe shows a significantly higher index across all temperatures. However, the correction at 300\,K is only slightly larger than that of SiC.
% This suggests that the correction is influenced not solely by electron-phonon or three-phonon interactions, but rather by the interplay between the two.
% Notably, the index itself primarily reflects the strength of three-phonon interactions entering in the form of dephasing effect.
%This already serves as a useful primary indicator for identifying materials worth further investigation.

\begin{figure}[t]
\centering
\includegraphics[width=\linewidth]{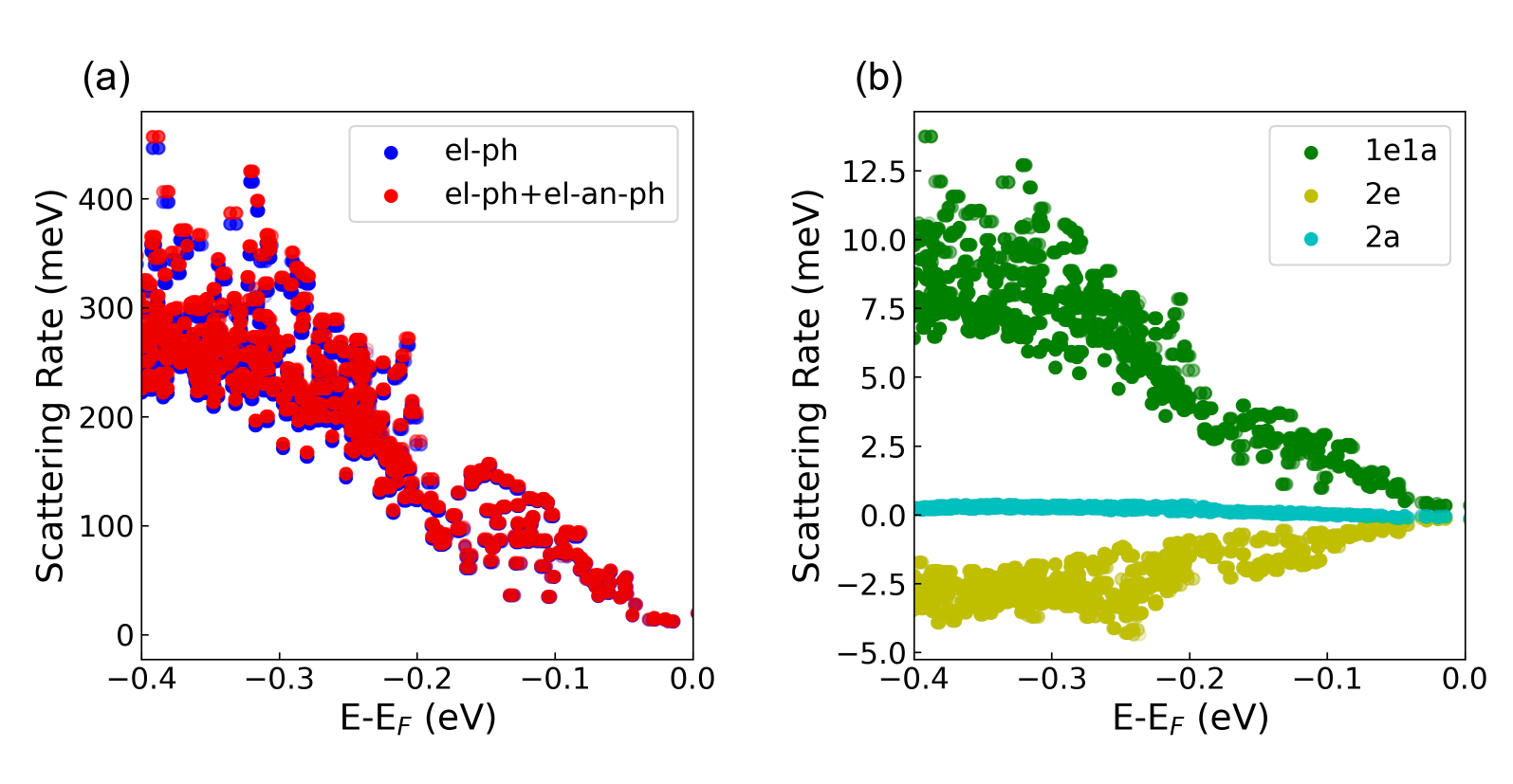}
\caption{(a) Electron-phonon scattering rate for PbTe with and without electron-anharmonic-phonon correction versus electron energy at $500$\,K; (b) electron-anharmonic-phonon correction versus electron energy at $500$\,K.}
\label{fig:index}
\end{figure}

Higher temperatures lead to an increase in the phonon population. To explore the potential impact of this on the anharmonic dephasing effects on electron-phonon coupling, we calculate the harmonic electron-phonon and anharmonic electron-phonon scattering rates of PbTe at $500$\,K, reported in Fig.\,\ref{fig:index}. These results should be compared with those at $300$\,K reported in Fig.\,\ref{fig:pbte}.
The electron–phonon scattering rates at $500$\,K in Fig.\,\ref{fig:index}(a) exhibit a pronounced increase compared to those at $300$\,K in Fig.\,\ref{fig:pbte}(d), as expected from the phonon population growth with increasing temperature. The total scattering rate at $500$\,K is still dominated by the harmonic electron-phonon coupling, but the anharmonic dephasing contribution to electron phonon coupling, depicted in Fig.\,\ref{fig:index}(b) at $500$\,K, also grows significantly compared to the results at $300$\,K in Fig.\,\ref{fig:pbte}(e). This is consistent with the expectation that increasing temperature enhances anharmonic corrections.

\subsection{Discussion}

Figures~\ref{fig:si}(e), \ref{fig:sic}(e), and~\ref{fig:pbte}(e) show that the different microscopic mechanisms contributing to anharmonic electron-phonon coupling in Si, SiC, and PbTe, can be positive or negative, an observation that can be rationalized by considering the scattering rate formulae in Eqs.\,(\ref{eq:gamma-1e1a})-(\ref{eq:gamma-2a}). 
According to Eq.\,(\ref{eq:gamma-1e1a}), the one-phonon absorption and one-phonon emission corrections are always positive. 
In contrast, the two-phonon absorption corrections are always negative, as indicated by Eq.\,(\ref{eq:gamma-2a}).
Finally, the two-phonon emission corrections, described by Eq.\,(\ref{eq:gamma-2e}), can be either positive or negative depending on the occupation number terms.

In this context, both Si and SiC exhibit a dominant negative contribution from 2e processes, while other processes are negligible. By contrast, PbTe exhibits a significant positive contribution from the 1e1a processes, but this contribution is partially cancelled by a negative contribution from the 2e processes, thereby reducing the total correction to the scattering rates.

More generally, different materials could exhibit different dominant microscopic scattering mechanisms, whose combination could lead to different strengths and signs for the overall anharmonic corrections. 
%processes can explain why these modifications are often neglected: self-cancellation occurs between the scattering contributions.

\subsection{Scaling and convergence}

\begin{figure*}[t]
\centering
\includegraphics[width=\linewidth]{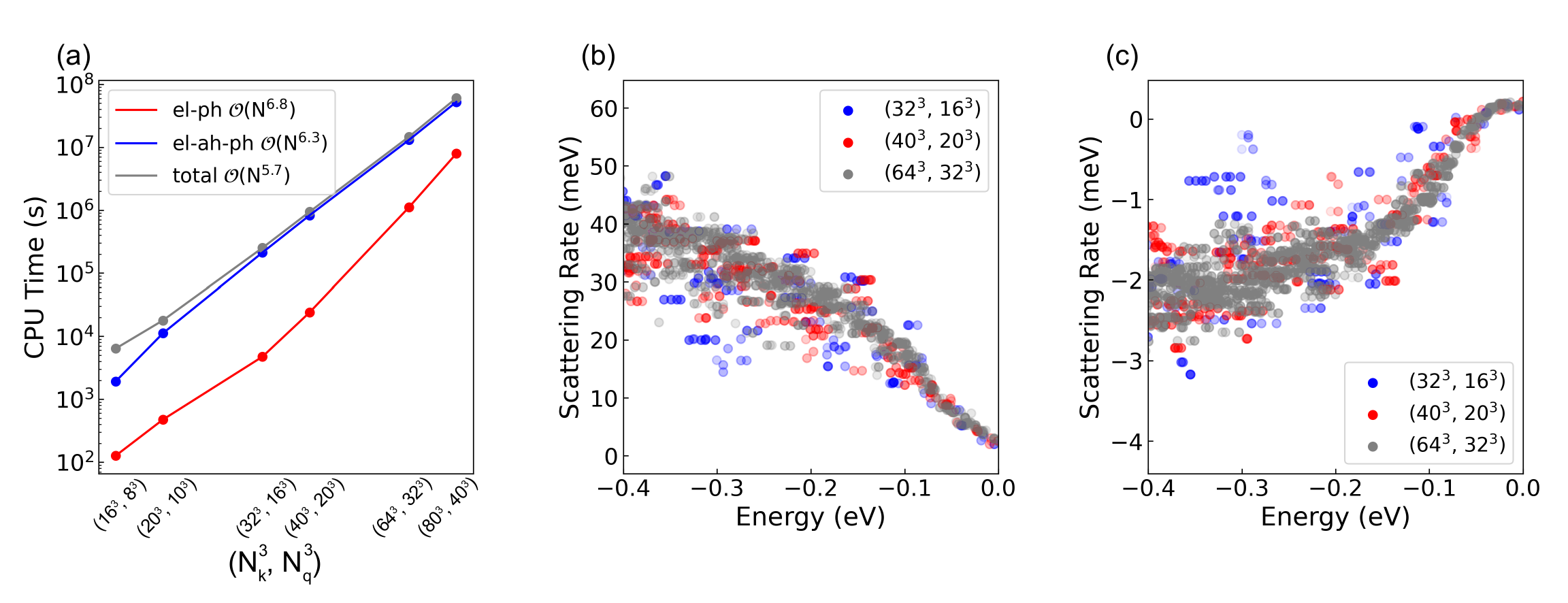}
\caption{(a) Scaling of the harmonic electron-phonon coupling calculation (red), electron–anharmonic-phonon coupling calculation (blue), and total calculation (grey) with respect to the electron and phonon reciprocal space grid sizes in Si (identical sizes in each direction for each grid; time measured in core-seconds); (b) electron-phonon scattering rates for different grid sizes; (c) electron-anharmonic-phonon coupling correction for different grid sizes.}
\label{fig:scale}
\end{figure*}

The results presented above show the computational feasibility of evaluating the role of anharmonic dephasing in electron-phonon coupling using fully first principles methods. Nonetheless, these calculations are computationally intensive, and in this section we discuss their scaling and convergence using the Si calculations as an example.

%The scaling and convergence behavior of our silicon calculations in silicon versus electron and phonon mesh densities is presented in Fig.\,\ref{fig:scale}.

The first principles evaluation of properties related to electrons and phonons requires the fine sampling of the associated Brillouin zones. In the examples discussed above, we use a uniform $\mathbf{k}$-point grid of size $N_{\mathbf{k}}\times N_{\mathbf{k}}\times N_{\mathbf{k}}$ to sample the electron Brillouin zone, and for simplicity we denote its size by $N_{\mathbf{k}}^3$. Similarly, we use a uniform $\mathbf{q}$-point grid of size $N_{\mathbf{q}}\times N_{\mathbf{q}}\times N_{\mathbf{q}}$ to sample the phonon Brillouin zone, and we denote its size by $N_{\mathbf{q}}^3$. 

To optimize the computational cost of our first principles implementation, we use a Fermi window to select the relevant states, inner-loop physical quantities are precomputed before entering nested loops, and tensor contraction is exploited for the fine-grained parallelization.
All reported calculations have been performed on CPU cores, specifically the Intel(R) Xeon(R) Gold 6248 @ 2.50GHz processors on the Young cluster of the Materials Modelling Hub of the UK. An implementation to GPU hardware will be reported elsewhere.

In Fig.\,\ref{fig:scale}(a), we plot the computational time as a function of the electron $\mathbf{k}$-point and phonon $\mathbf{q}$-point reciprocal space grid sizes. 
The observed overall scaling for the anharmonic electron-phonon coupling calculations follows an approximate scaling of $\mathcal{O}(N^{5.7})$. Since the electron and phonon grid sizes increase simultaneously, this corresponds to an approximate scaling of $\mathcal{O}(N^{3})$ for each grid individually.
%The results also imply that the scaling is $\mathcal{O}(N)$ with respect to the grid size along each direction.
Figure~\ref{fig:scale}(a) also shows that the scaling of electron-anharmonic-phonon calculation is similar to that of the harmonic electron-phonon calculation, but with a much larger prefactor.

Figure~\ref{fig:scale}(b) shows the electron–phonon scattering rates calculated within the harmonic approximation as a function of electron energy, and Fig.\,\ref{fig:scale}(c) shows the anharmonic contribution to electron-phonon scattering rates as a function of electron energy. 
In both cases, the results are shown for different reciprocal space sampling grid sizes, illustrating the convergence of the results with respect to both electron and phonon grid sizes. 
The results indicate that the scattering rates converge with increasing grid size, as evidenced by the overlap of the data for (40$^3$, 20$^3$) (red) and (64$^3$, 32$^3$) (grey) grids in both cases. %This suggests that a 40$-$20 grid already provides reasonably accurate results, with only minor differences from the denser 64$-$32 sampling that might be led by the same smearing but denser sampling.
We highlight that these results only illustrate the general convergence pattern of the scattering rates. Adequate convergence calculations should be performed for the specific physical quantities to be computed as derived from these scattering rates, as done for example for the electron conductivity of MgB$_2$ in the companion work.

\section{Conclusions}

We have presented a systematic discussion of electron-phonon coupling in anharmonic crystals. Focusing on the finite lifetimes of phonons in anharmonic crystals, we have derived an expression for the lowest-order anharmonic dephasing in the electron-phonon interaction, which brings together electron-phonon and phonon-phonon scattering mechanisms. We have also presented a first principles implementation of anharmonic dephasing in electron-phonon coupling, and explored its impact on the electron-phonon scattering rates in silicon, silicon carbide, and lead telluride. In all these compounds, the effects of anharmonic dephasing are present but small. In the companion work, we present calculations of the effects of anharmonic dephasing on electron-phonon coupling in magnesium diboride, finding that they dramatically affect the electron conductivity in this compound. 
%Starting from the fundamental free-electron and free-phonon system and extending to the final electron–anharmonic phonon coupling, we are expecting to offer a clear framework for analyzing such problems by systematically decomposing and classifying the underlying physical mechanisms, along with deep insights into the complex system where electrons, phonons, and their intra- and inter-interactions play a central role.

% When the scale of investigation becomes smaller and precision grows, coherence effects and dephasing phenomena become increasingly significant, playing a crucial role in determining the behavior of quantum systems.
More generally, our work contributes to an ongoing effort to expand the standard framework to evaluate electron-phonon interactions in materials from first principles. Our work is the first to show the importance of phonon lifetimes in the electron-phonon interaction, and provides a solid foundation to explore this regime in different materials.
%As the scale of investigation decreases and experimental precision improves\,\cite{Liu2024}, previously negligible quantum effects emerge, challenging conventional theoretical frameworks.
%Such observations may necessitate a reevaluation of a growing range of problems with dephasing happening during coupling, which may result in minor corrections — or, more compellingly, reveal new physics and provide answers to previously unresolved phenomena, all rooted in the remarkable diversity of the condensed matter world.

Moving forward, it would be interesting to establish a more comprehensive framework to systematically investigate electron-phonon coupling in systems with anharmonic phonon behavior. An important future direction is the incorporation of all relevant corrections, including higher-order contributions, something that could potentially be accomplished using functional derivative techniques. %While such anharmonic correction effects are often regarded to be small, they may be significant in certain systems. Our current work serves as a proof of principle, and we leave the development of these more rigorous and computationally demanding approaches for future study.
Alongside developments on the theoretical front, we will describe elsewhere the \texttt{DaoQuantum} code used in the calculations reported here. We expect it to become a highly customizable and high-performance tool, enabling extensive studies accessible to the broader research community.

\begin{acknowledgements}
M.K. and B.M. acknowledge financial support from the Gianna Angelopoulos Programme for Science, Technology, and Innovation. B.M. also acknowledges support from a UKRI Future Leaders Fellowship [MR/V023926/1].
Computational support was provided by the UK Materials and Molecular Modelling Hub, which is partially funded by EPSRC [EP/P020194], and for which access was obtained via the UKCP consortium and funded by EPSRC grant [EP/X035891/1].
%The computational resources were provided by the Cambridge Tier-2 system operated by the University of Cambridge Research Computing Service and funded by EPSRC [EP/P020259/1] and by the UK National Supercomputing Service ARCHER2, for which access was obtained via the UKCP consortium and funded by EPSRC [EP/X035891/1].
\end{acknowledgements}

\bibliography{ref} % Include references from ref.bib
\end{document}